\documentclass[fleqn,usenatbib]{mnras}
\usepackage{newtxtext,newtxmath}
\usepackage[T1]{fontenc}

\DeclareRobustCommand{\VAN}[3]{#2}
\let\VANthebibliography\thebibliography
\def\thebibliography{\DeclareRobustCommand{\VAN}[3]{##3}\VANthebibliography}

\usepackage{graphicx}	
\usepackage{amsmath}	
\usepackage{tikz}
\usetikzlibrary{positioning, arrows.meta, shapes, calc}
\usepackage{array}  
\usepackage{caption}  
\usepackage{fix-cm}
\usepackage{bm}
\usepackage{graphicx}
\usepackage{booktabs}
\usepackage{boldline}
\usepackage{comment}
\usepackage{xcolor}

\newcommand\Lya{Ly$\alpha$} 
\newcommand\HI{\hbox{H$\,\rm \scriptstyle I$}}
\newcommand\HII{\hbox{H$\,\rm \scriptstyle II$}}

\title[Inference from 21-cm forest]{Efficient neutral-IGM inference from noisy 21-cm forest spectra with latent-space U-Net encoding and XGBoost}

\author[Patil et al.]{
Sameer K. Patil,$^{1}$\thanks{E-mail: sameerkpatil@gmail.com}
Tomáš Šoltinský$^{2,3,4},$
Soumak Maitra$^{1}$
and Girish Kulkarni$^{1}$
\\
$^{1}$Tata Institute of Fundamental Research, Homi Bhabha Road, Mumbai 400005, India\\
$^{2}$INAF–Osservatorio Astronomico di Trieste, Via G.B. Tiepolo, 11, I-34143 Trieste, Italy\\
$^{3}$IFPU, Institute for Fundamental Physics of the Universe, Via Beirut 2, I-34151 Trieste, Italy\\
$^{4}$INFN, Sezione di Trieste, Via Valerio 2, I-34127 Trieste, Italy\\
}

\date{Accepted ---. Received ---; in original form ---}
\pubyear{\the\year{}}

\begin{document}
\label{firstpage}
\pagerange{\pageref{firstpage}--\pageref{lastpage}}
\maketitle

\begin{abstract}
The 21-cm forest, comprising narrow absorption features imprinted on the radio spectra of high-redshift radio-loud quasars by intervening neutral hydrogen, offers a uniquely sensitive probe of the thermal state of the neutral intergalactic medium (IGM) during the epoch of reionization. Although over 30 such quasars are now known at $z > 5.5$, the signal remains elusive in practice, owing to instrumental noise, the intrinsic weakness of the absorption features, and the limited brightness of available background sources. Recent studies have focused on the one-dimensional transmission power spectrum as a statistical observable, but this approach also demands high signal-to-noise ratios. Here, we present a systematic comparison of five inference pipelines for recovering IGM parameters from mock 21-cm forest spectra at $z = 6$, incorporating realistic instrumental noise and telescope characteristics. We show that likelihood-free inference based on machine learning substantially outperforms traditional Bayesian methods. In particular, our most effective method dispenses with the power spectrum entirely: we use a convolutional U-Net to extract a latent-space encoding of the input spectrum and perform parameter regression using XGBoost. This approach yields accurate constraints on the IGM neutral fraction and X-ray heating efficiency even with a single 50-hour uGMRT sightline, which is an orders-of-magnitude improvement in integration time relative to existing techniques. We publicly release our code, training data, and models. Beyond the 21-cm forest, these results underscore the promise of hybrid deep learning and gradient-boosted inference techniques for extracting physical information from low-SNR data across astrophysics.
\end{abstract}


\section{Introduction}
Observations of the redshifted 21-cm line of neutral hydrogen in the intergalactic medium (IGM) offer a powerful window into the early universe, particularly the epochs of cosmic dawn and reionization. This potential has spurred a broad spectrum of experimental efforts, including measurements of the sky-averaged (``global'') \mbox{21-cm} signal as well as of its spatial fluctuations. These approaches, which use the cosmic microwave background (CMB) as a backlight, have dominated the field thus far. However, an especially promising and complementary probe is the 21-cm forest---a series of narrow absorption features imprinted on the radio spectra of high-redshift background sources by intervening patches of neutral hydrogen in the IGM \citep{2002ApJ...577...22C,Furlanetto_2002,2006PhR...433..181F,Ciardi_2013,Soltinsky_2021,Soltinsky_2023}. Conceptually analogous to the Ly$\alpha$ forest observed in the optical and ultraviolet spectra of quasars, the 21-cm forest, seen in the radio, arises from the hyperfine transition of neutral hydrogen and serves as a uniquely sensitive tracer of small-scale structure in the neutral IGM during the epoch of reionization \citep{Xu_2009,Mack_2012,Soltinsky_2021,Shao_2023}. Crucially, it may represent one of the few practical means of accessing the cold neutral IGM during this epoch. (The only other probe so far studied for this is the Mg\,\textsc{ii} line forest \citep{2021MNRAS.506.2963H, 2024MNRAS.535..223T}, which relies on the IGM being sufficiently chemically enriched.) as this phase is effectively opaque to Ly$\alpha$ photons \citep{Soltinsky_2025}. Interest in the 21-cm forest has intensified in recent years, driven by the discovery of an increasing number of high-redshift, radio-loud quasars,\footnote{There are 34 radio-loud quasars known at $z \gtrsim 5.5$ at the time of writing. We maintain an up-to-date list at \url{https://tomassoltinsky.github.io/eor}.} and by a growing consensus that reionization concluded later than previously believed \citep{2019MNRAS.485L..24K,2022MNRAS.514...55B}, which enhance the probability of encountering sightlines intersecting substantial reservoirs of cold neutral gas. Current facilities such as the upgraded Giant Metrewave Radio Telescope (uGMRT) and the Low Frequency Array (LOFAR) are already capable of probing the 21-cm forest at redshifts $z \gtrsim 6$ \citep{Wayth_2015, Gupta_2017,Shimwell_2017, Kondapally_2021}, while the forthcoming Square Kilometre Array (SKA) promises to dramatically extend the reach and sensitivity of such studies \citep{Soltinsky_2025,Koopmans_2015, Braun_2019,Mellema_2015, Ciardi_2015}. 

The 21-cm forest offers a uniquely versatile probe of the high-redshift universe, sensitive to a broad array of cosmological and astrophysical processes. The strength and statistics of the forest directly depend on the IGM ionization fraction and thermal state: cold, neutral gas yields prominent absorption features, while X-ray or shock-heated regions suppress the signal \citep{Xu_2009,Mack_2012,Soltinsky_2021}. This makes the 21-cm forest a powerful tracer of both the timing of reionization and the history of IGM heating, complementing other EoR observables. Moreover, its sensitivity to small-scale structure enables constraints on the microphysical properties of dark matter, as warm or otherwise non-cold dark matter models that suppress the small-scale matter power spectrum, thereby diminishing the richness in the 21-cm forest \citep{2014PhRvD..90h3003S,2015MNRAS.451..467S,Shao_2023,2023PASJ...75S..33V}. Similarly, massive neutrinos imprint their free-streaming scale on the early matter distribution, altering the prevalence of absorption lines and allowing the 21-cm forest to probe the sum of neutrino masses \citep{2014PhRvD..90h3003S,Shao_2025}. Finally, by requiring bright background radio sources to illuminate the intervening IGM, forest observations inherently constrain the abundance, luminosity, and lifetime of high-redshift radio-loud quasars, thus providing indirect insights into the growth of early supermassive black holes \citep{ewallwice2014,Soltinsky_2023}. Together, these connections make the 21-cm forest a rich observable for simultaneously exploring reionization, IGM thermal history, dark matter physics, neutrino masses, and the population of early luminous black holes.

Despite these inducements, extracting accurate and reliable cosmological information from 21-cm forest observations is fraught with technical challenges. The primary challenge is the intrinsic faintness of the 21-cm absorption lines relative to the instrumental thermal noise. 
\citet{Soltinsky_2025} found that for a simulated uGMRT over $500\,\rm hr$ observed spectrum, frequency channels with high signal-to-noise ratio (SNR) are $>20$ times less abundant than the ones at more representative values of $\rm{SNR}\sim0.4$.
These authors also reported a $p$-value of $0.337 \pm{0.228}$ on a two-sided Kolmogorov-Smirnov test for the observed and noise distribution on mock spectra for uGMRT over $50\,\rm hr$. This $p$-value is very large and implies that the observed distribution function is well within what might be expected due to random sampling variability. 
So the direct detection of this absorption requires very bright ($\gtrsim 10$--$100~\mathrm{mJy}$) background sources at high redshifts ($z\gtrsim  8$), which are evidently rare, or very long integration times, or instruments of very high sensitivity. 
To address these challenges, \citet{2020ApJ...899...16T} proposed a statistical approach - one-dimensional (1D) power spectra of the 21-cm forest spectra of multiple fainter background objects ($\sim 1$-$10~\mathrm{mJy}$), which are likely to be more abundant.
Building on this framework, \citet{Soltinsky_2025} developed inference procedures based on likelihood-based Bayesian methods to constrain the thermal and ionization state of the neutral IGM during the epoch of reionization. 
They showed that it is possible to detect the 1D power spectrum of the 21-cm forest spectrum, especially at large scales of $k \lesssim8.5~\mathrm{MHz}^{-1}$ with $500\,\rm hr$ of uGMRT time and at $k\lesssim32.4~\mathrm{MHz}^{-1}$ with  SKA1-low over $50\,\rm hr$ if the intergalactic medium (IGM) is 25\% neutral and the neutral hydrogen regions have a spin temperature of $\lesssim30\,\rm{K}$. 
They also inferred that a null detection of the signal can be translated into constraints which are tighter than the currently available measurements. 
Moreover, they argued that a null detection of the 1D 21-cm forest power spectrum with $50\,\rm hr$ of the uGMRT observations can be competitive with other observations in disfavouring models of significantly neutral and cold IGM at $z\sim6$.

The high integration time demands of previous studies, coupled with their limited accuracy in the weak-signal regime of the 21-cm forest, motivate the search for more effective parameter inference techniques. In particular, the Bayesian Markov Chain Monte Carlo (MCMC) approach based on the 1D transmission power spectrum, though well-suited to Gaussian-distributed signals, exhibits notable shortcomings. It fails to capture non-Gaussian statistical features that are intrinsic to cosmological signals \citep{2025arXiv250414656S,Shaw_2020,Wolfson_2023}, highlighting the need for more sophisticated feature extraction strategies. Furthermore, the presence of instrumental thermal noise degrades inference accuracy, underscoring the importance of robust noise mitigation.

\begin{figure*}
\centering
\resizebox{0.95\textwidth}{!}{
\begin{tikzpicture}[
    font=\small,
    >=Stealth,
    startend/.style={rectangle,draw,fill=gray!25,rounded corners,
                     minimum width=33mm,minimum height=12pt,align=center},
    ps/.style   ={rectangle,draw,fill=cyan!20,rounded corners,
                 minimum width=33mm,minimum height=10mm,align=center},
    noise/.style={rectangle,draw,fill=yellow!35,rounded corners,
                 minimum width=33mm,minimum height=10mm,align=center},
    unet/.style ={rectangle,draw,fill=violet!25,rounded corners,
                 minimum width=33mm,minimum height=10mm,align=center},
    latent/.style={rectangle,draw,fill=violet!40,rounded corners,
                 minimum width=33mm,minimum height=10mm,align=center},
    mcmc/.style ={rectangle,draw,fill=orange!35,rounded corners,
                 minimum width=33mm,minimum height=10mm,align=center},
    xgb/.style  ={rectangle,draw,fill=blue!30,rounded corners,
                 minimum width=33mm,minimum height=10mm,align=center},
    col/.style  ={anchor=north},
    myarrow/.style={->,thick,rounded corners}
]

\def\xsep{38mm}   
\def\ysep{17mm}   

\foreach \i in {0,...,4}{
   \coordinate (X\i) at (\i*\xsep,0);
}

\coordinate (Xc) at (X2);

\node[startend] (start) at ($(Xc)+(0,20mm)$)
      {\strut Noisy 21-cm\\forest spectrum\\$F_{21}^{S+N}$};

\node[startend] (end) at ($(Xc)+(0,-100mm)$)
      {\strut $\bm{\theta}$\\$[\langle x_{\rm HI}\rangle, \log_{10}f_X$]};

\newcommand{\placenode}[5]{%
  \node[#3, col] (#4) at ($(X#1)+(0,-#2*\ysep)$) {#5};
}


\placenode{0}{1}{ps}{ps1}{1D PS}
\placenode{0}{3}{mcmc}{mcmc1}{Bayesian\\MCMC}

\placenode{1}{1}{ps}{ps2}{1D PS}
\placenode{1}{2}{noise}{sub2}{Noise\\subtraction}
\placenode{1}{3}{mcmc}{mcmc2}{Bayesian\\MCMC}

\placenode{2}{1}{ps}{ps3}{1D PS}
\placenode{2}{3}{xgb}{xgb3}{XGBoost}

\placenode{3}{1}{unet}{unet4}{U-Net\\denoising}

\placenode{3}{2}{ps}{ps4}{1D PS}
\placenode{3}{3}{xgb}{xgb4}{XGBoost}

\placenode{4}{1}{unet}{enc5}{U-Net\\encoder}
\placenode{4}{2}{latent}{lat5}{Latent features}
\placenode{4}{3}{xgb}{xgb5}{XGBoost}

\foreach \n in {ps1,ps2,ps3,unet4,enc5}{
   \draw[myarrow] (start.south) to[out=-90,in=90] (\n.north);
}

\draw[myarrow] (ps1) -- (mcmc1);
\draw[myarrow] (ps2) -- (sub2);
\draw[myarrow] (sub2) -- (mcmc2);
\draw[myarrow] (ps3) -- (xgb3);
\draw[myarrow] (unet4) -- (ps4);
\draw[myarrow] (ps4) -- (xgb4);
\draw[myarrow] (enc5) -- (lat5);
\draw[myarrow] (lat5) -- (xgb5);

\foreach \n in {mcmc1,mcmc2,xgb3,xgb4,xgb5}{
   \draw[myarrow] (\n.south) to[out=-90,in=90] (end.north);
}
\node[font=\bfseries, fill=white, fill opacity=1, text opacity=1] at ($(X0)+(0,-10mm)$) {Method~A1};
\node[font=\bfseries, fill=white, fill opacity=1, text opacity=1] at ($(X1)+(0,-10mm)$) {Method~A2};
\node[font=\bfseries, fill=white, fill opacity=1, text opacity=1] at ($(X2)+(0,-10mm)$) {Method~B1};
\node[font=\bfseries, fill=white, fill opacity=1, text opacity=1] at ($(X3)+(0,-10mm)$) {Method~B2};
\node[font=\bfseries, fill=white, fill opacity=1, text opacity=1] at ($(X4)+(0,-10mm)$) {Method~B3};

\end{tikzpicture}}
\caption{Schematic data flow of the five parameter inference pipelines explored in this work. The diagram traces the transformation from the input noisy 21-cm forest spectrum (top) to the final parameter estimates $\bm{\theta}$ (bottom). Methods A1 and A2 use likelihood-based Bayesian methods, while Methods B1, B2, and B3 employ likelihood-free machine learning techniques.}
\label{fig:five-inference-pipelines}
\end{figure*}
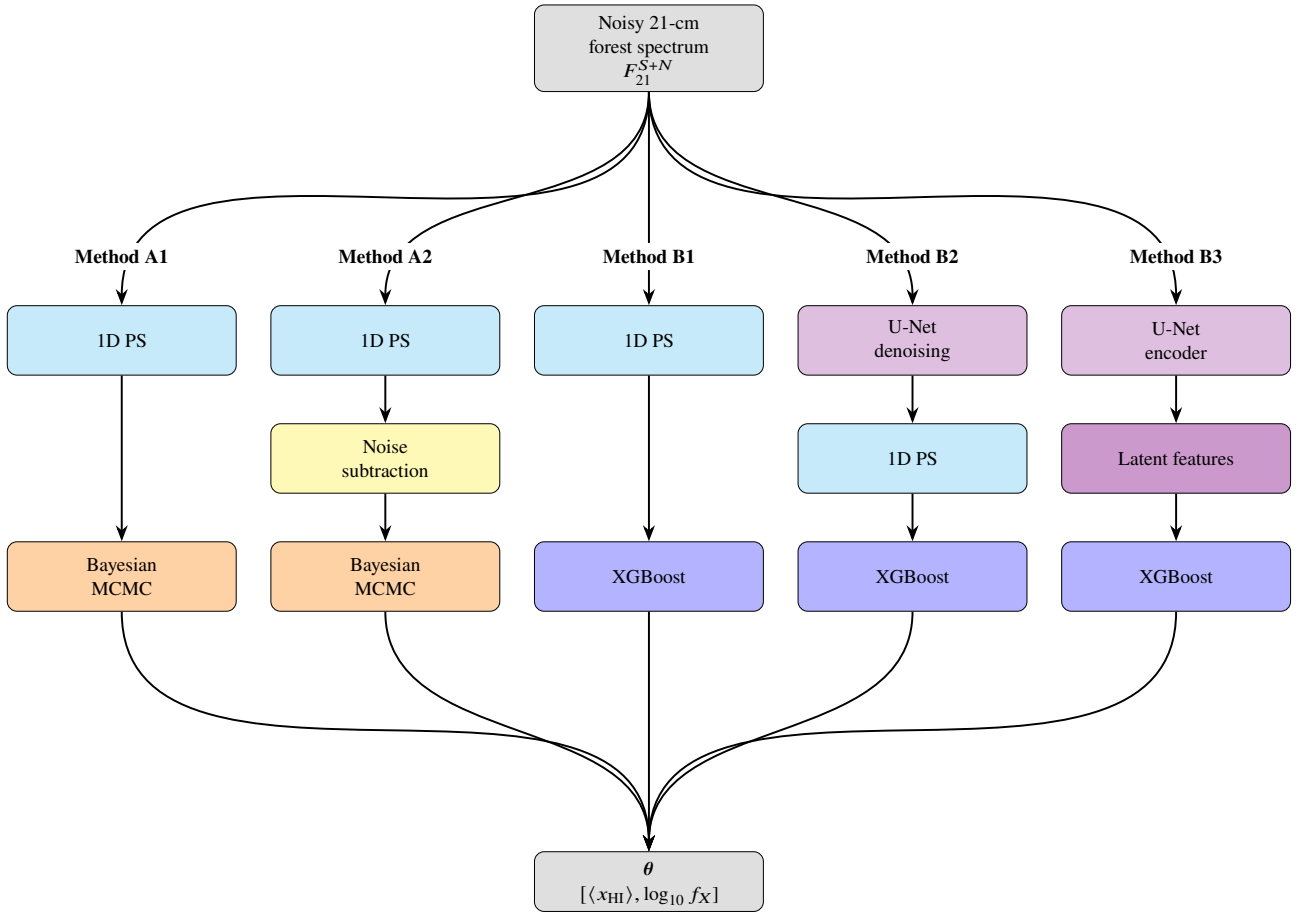

In this paper, we systematically evaluate five distinct pipelines for inferring IGM parameters from noisy 21-cm forest spectra. Our methodology involves generating mock observations and applying each inference technique in turn. Two of these approaches are conventional, likelihood-based Bayesian methods. The remaining three employ deep learning techniques, specifically the U-Net architecture and the XGBoost algorithm. While some methods make use of summary statistics such as the 1D transmission power spectrum, we also investigate denoising strategies, either via explicit noise subtraction or through learned denoising with neural networks. Finally, we explore an approach that bypasses both denoising and the power spectrum, instead relying on latent-space encodings extracted from the raw spectrum. The five methods are illustrated schematically in Fig.~\ref{fig:five-inference-pipelines}. Through a quantitative comparison, we identify a novel inference framework that outperforms the others across a wide range of conditions.

Recent advances in machine learning (ML), particularly in deep learning, offer promising tools for denoising astrophysical data. Convolutional neural networks (CNNs), with their strong feature extraction capabilities, have proven adept at identifying faint astrophysical signals embedded in noise. They have been applied successfully to a variety of problems, including the detection and characterisation of Ly$\alpha$ absorption in quasar spectra \citep{Parks_2018} and the inference of cosmological parameters directly from Ly$\alpha$ forest flux \citep{Huang_2021}. One especially relevant architecture is the denoising autoencoder (DAE), which learns to reconstruct clean inputs from corrupted data by extracting robust features \citep{dae}.

Among DAE variants, the U-Net architecture—originally developed for biomedical image segmentation \citep{ronneberger2015unetconvolutionalnetworksbiomedical}—has shown particular promise in astrophysical contexts. It excels at preserving local spectral features while suppressing noise, thanks to its use of skip connections between encoder and decoder layers. U-Nets have been applied to 21-cm intensity mapping foreground subtraction \citep{Makinen_2021}, radio-frequency interference mitigation, recovery of diffuse radio emission \citep{Gheller_2021}, and denoising of optical images \citep{Vojtekova_2020}. In such applications, the network’s latent encoding learns to isolate signal structure from stochastic noise.

In this study, we develop a new framework that combines U-Net denoising with gradient-boosted regression for parameter inference from 1D 21-cm forest spectra. The key idea is to use the CNN to compress the noisy input into a latent-space representation that preserves salient astrophysical features. Our models are trained on mock datasets incorporating both realistic IGM absorption signatures and instrumental noise. Of the architectures explored, U-Net consistently yielded the best performance in terms of convergence and reconstruction fidelity. CNN-based approaches are particularly well suited to this problem, as they can learn to capture complex, non-Gaussian features that traditional power spectrum–based methods often miss \citep{Gillet2019}.

To infer physical parameters from the latent features extracted by the U-Net, we use XGBoost (eXtreme Gradient Boosting), a fast and scalable ensemble learning method based on gradient-boosted decision trees \citep{Chen2016}. XGBoost offers several advantages: it handles complex non-linear mappings effectively, requires minimal tuning compared to deep neural networks, and provides interpretable outputs through feature importance metrics. Its architecture includes regularisation, sparsity-aware algorithms, and parallel computation, making it well suited to structured data regression tasks.  XGBoost has already demonstrated its versatility across a range of astrophysical applications. For example, it has been used to infer halo properties from simulations \citep{Zhao_2025}, to identify brightest cluster galaxies in large surveys \citep{Tian_2025}, and to analyse spectro-photometric datasets \citep{anders2023parameters300million}. Drawing on these successes, we take a further step, and apply XGBoost to regress IGM parameters from the latent features derived via U-Net encoding, with the aim of quantifying the achievable accuracy in the presence of noise.  This method foregoes the 1D power spectrum as well as denoising entirely, instead relying on the U-Net to extract relevant features directly from the noisy input spectrum.

Our broader objective is to evaluate whether such hybrid deep-learning and tree-based approaches can relax telescope integration time requirements, thereby enabling more efficient constraints on the thermal and ionisation state of the IGM from 21-cm forest observations during the epoch of reionization. By leveraging the strengths of neural networks for feature extraction and gradient-boosted trees for regression, we aim to demonstrate that meaningful astrophysical inference is possible even in low signal-to-noise regimes. This could significantly expand the scientific reach of current and upcoming radio facilities, not only making the 21-cm forest a practical probe in IGM studies, but also potentially making available a technique for extracting physical information from other low-SNR astrophysical datasets.

This paper is organised as follows. Section~\ref{sec:simulation} details our methodology for generating realistic simulated datasets of the 21-cm forest, including the underlying astrophysical assumptions and the modelling of observational noise. The remainder of the paper examines five distinct inference pipelines, designated as A1, A2, B1, B2, and B3, which are schematically illustrated in Fig.~\ref{fig:five-inference-pipelines}. In Section~\ref{sec:MCMC_noisesubtracted}, we present methods~A1 and A2, which are based on conventional likelihood-based approaches combining power spectrum analysis with Bayesian inference. Section~\ref{sec:Inference_ML} introduces Methods~B1, B2, and B3, which use ML techniques for a likelihood-free parameter regression. In Section~\ref{sec:comparison}, we provide a comparative analysis of the performance of all five methods. We conclude in Section~\ref{sec:conclusion} by summarising our key findings and discussing their implications for future observational strategies.

\section{Simulating the 21-cm forest signal}\label{sec:simulation}

Our model for the 21-cm forest signal in the epoch of reionization is identical to that used in \citet{Soltinsky_2025}. We direct the reader to that work for a detailed description of the simulation and signal forward-modelling methodology. In this section, we briefly summarize the main aspects of the modelling relevant to our analysis.

\subsection{The IGM model}\label{sec:IGM_model}

\begin{figure}
    \centering
    \includegraphics[width=1\linewidth]{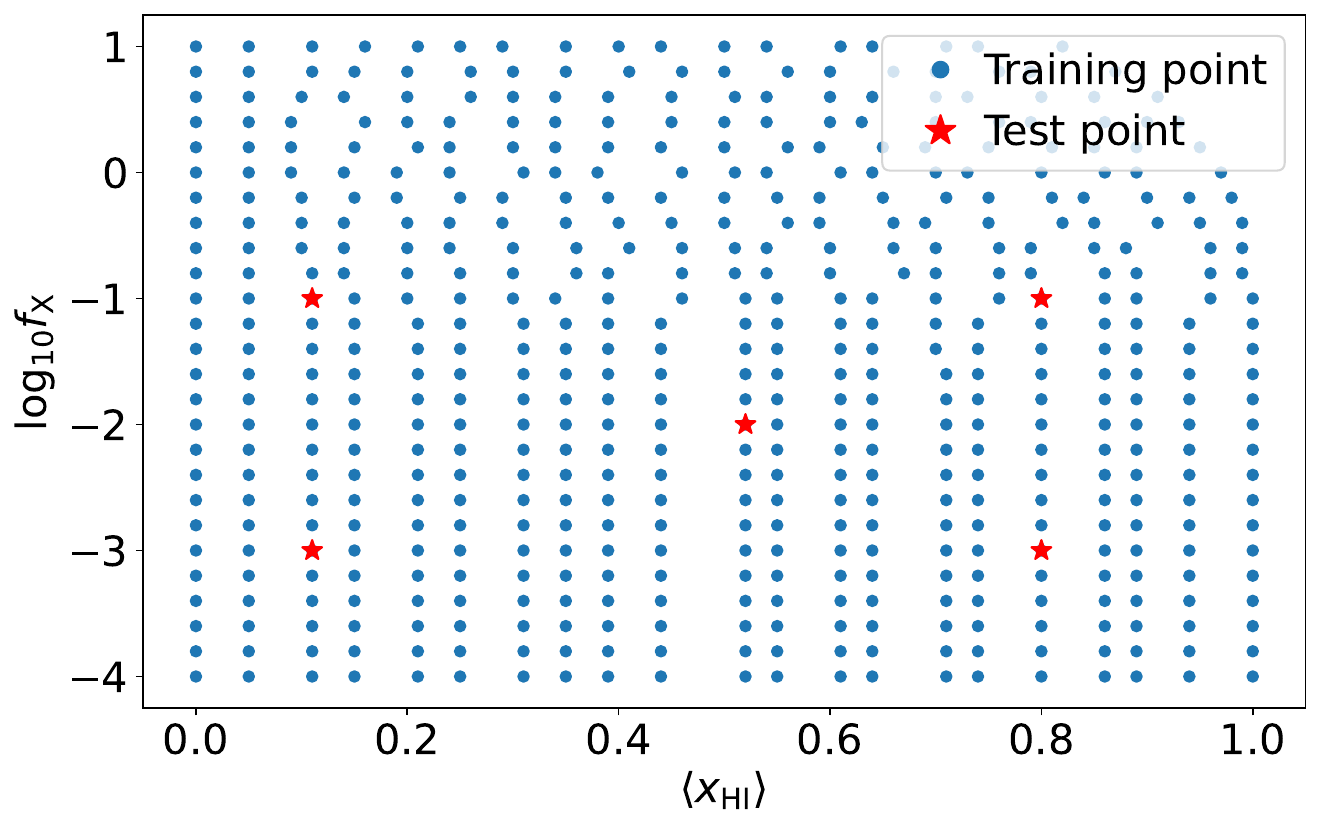}
    \caption{Sampling of the parameter space for 21-cm forest simulations. The model is exactly evaluated at all 534 points shown here. Of these, blue circles indicate the 529 parameter combinations used for training the machine learning models, while red stars mark the 5 test points reserved for evaluating inference accuracy. The test points are \{$\langle x_{\mathrm{HI}}\rangle$, $\mathrm{log}_{10}f_{\rm X}$\}: [\{0.11, $-3.0$\}, \{0.11, $-1.0$\}, \{0.52, $-2.0$\}, \{0.80, $-1.0$\}, \{0.80, $-3.0$\}].}
    \label{fig:training-test-points}
\end{figure}

To model the IGM at $z=6$ we utilize the publicly available semi-numerical code \textsc{21cmFAST}\footnote{\hyperlink{https://github.com/21cmFAST/21cmFAST}{https://github.com/21cmFAST/21cmFAST}} version 3.3.1 \citep{Mesinger_2011_21CMFAST,Murray_2020}. The simulations are based on a flat $\Lambda$CDM cosmology assuming $\Omega_{\Lambda}=0.692$, $\Omega_{\rm m}=0.308$, $\Omega_{\rm b}=0.0482$, $\sigma_8=0.829$, $n_{\rm s}=0.961$, $h=0.678$ \citep{planck2014} and a primordial helium fraction by mass of $Y_{\rm p}=0.24$ \citep{Hsyu_2020}. The simulations are run in a volume of $\left(50\, \rm cMpc\right)^3$ resolving $\sim195\,\rm ckpc$ large cells (i.e. $256^3$ cells in total).

We are interested in varying the thermal and ionization state of the cold and neutral IGM. The thermal state in the neutral regions of the intergalactic gas is dictated by the X-ray background radiation. The X-ray background radiation luminosity is parametrized here as
\begin{equation}
L_{\rm X}=3.4\times 10^{40}\rm\,erg\,s^{-1} \, \mathnormal{f_{\rm X}} \left(\frac{SFR}{1\,M_{\odot}\rm\,yr^{-1}}\right), \label{eq:LX} 
\end{equation}
\noindent
where the SFR is the star formation rate and $f_{\rm X}$ is the X-ray background radiation efficiency \citep{Furlanetto_2006b}. To follow this parametrization in the simulations we set the simulation parameter \textsc{L\_X} to $\mathrm{log}_{10}\left(3.4\times 10^{40}\rm\,erg\,s^{-1}\mathnormal{f_{\rm X}}\right)$ and vary the $f_{\rm X}$. Note that while there are constraints on the IGM preheating from the Murchison Widefield Array (MWA), Low-Frequency Array (LOFAR) and Hydrogen Epoch of Reionization Array (HERA) at $z>6.5$ \citep[][respectively]{Greig_2021_MWA,Greig_2021_LOFAR,Hera_2023}, especially disfavouring the models with no IGM preheating, the $f_{\rm X}$ is still largely unconstrained.
We note that there has been a recent measurement of $f_{\rm X}=0.5^{+6.3}_{-0.3}$ by \citet{Dhandha_2025} but this depends strongly on the assumed model. In this work the $\mathrm{log}_{10}f_{\rm X}$ is varied between $-4$ and $1$.

We characterise the ionization state of the IGM using the mean neutral hydrogen fraction, $\langle x_{\rm HI} \rangle$. We modulate this quantity by varying the ionizing efficiency of high-redshift galaxies in \textsc{21cmFAST}, which determines the total number of ionizing photons introduced into the IGM. 
The mean free path of these photons is held fixed at $0.75\,\rm pMpc$, consistent with the measurement at $z = 6$ by \citet{Becker_2021}. 
This choice sets the typical distance ionizing photons can travel before being absorbed and consequently defines the characteristic size of \HII\ regions. 
Within the \textsc{21cmFAST} simulations, these two quantities are governed by the parameters \textsc{HII\_EFF\_FACTOR} and \textsc{R\_BUBBLE\_MAX}, respectively. However, note that only the former one is varied.
We explore the full physically meaningful range of $\langle x_{\rm HI} \rangle$, spanning $[0, 1]$. 

The parameter combinations used in this work are shown in Fig.~\ref{fig:training-test-points}, where we sample the parameter space at 534 distinct points. 
Note that the top right corner of the parameter space corresponding to high $\langle x_{\rm HI} \rangle$ and $\mathrm{log}_{10}f_{\rm X}$ is not populated. This is due to the fact that in our models such strong X-ray background ionizes the IGM significantly even when the ionizing efficiency of high-redshift galaxies is set to zero.
These 534 points serve as the basis for two complementary inference approaches. 
For likelihood-based inference, we interpolate the likelihood over the parameter space using these computed models. 
For likelihood-free inference, we designate 529 of these points for training our machine learning algorithms, reserving 5 distinct models for testing. 
In Fig.~\ref{fig:training-test-points}, training points are marked in blue and test points are indicated by red stars. 
In total, our analysis draws upon 534 distinct IGM realisations. From each simulation, we extract 1000 lines of sight (LoS) of the underlying fields: gas overdensity $\Delta$, peculiar velocity $v_{\rm pec}$, neutral hydrogen fraction $x_{\rm HI}$, and gas kinetic temperature $T_{\rm K}$.  Each sightline has 2762 pixels corresponding to a 8~kHZ spectral resolution.

\subsection{The 21-cm forest}
\label{sec:21cm_forest}

Once the LoS data with various fields describing the IGM properties is constructed we can compute the optical depth to the 21-cm photons in a discrete form as \citep[e.g.][]{Furlanetto_2002}
\begin{align}
\tau_{\rm 21, i} =~& \frac{3h_{\rm p}c^{3}A_{10} }{32\pi^{3/2}\nu_{21}^{2}k_{\rm B}} \frac{\delta v}{H(z)} \nonumber \\
             & \times \sum_{j=1}^{N}\frac{n_{{\rm HI}, j}}{b_{j}T_{{\rm S}, j}}\exp\left( - \frac{ (v_{{\rm H},i}-u_{j})^{2}} {b^{2}_{j}}\right), \label{eq:tau21_discrete}
\end{align}
\noindent
where $i$ is the pixel of interest, $h_{\rm p}$ is the Planck constant, $k_{\rm B}$ is the Boltzmann constant, $c$ is the speed of light, $A_{10}=2.85\times 10^{-15}\rm\,s^{-1}$ is the Einstein spontaneous emission coefficient for the spin-flip transition, $\delta v$ is the velocity width of the pixels, $H(z)$ is the Hubble parameter, $n_{\rm HI}=\Delta\rho_{\rm c}x_{\rm HI}$ is the \HI~number density, $\rho_{\rm c}$ is the critical density of the universe, $b=(2k_{\rm B}T_{\rm K}/m_{\rm H})^{1/2}$ is the Doppler parameter, and $m_{\rm H}$ is the hydrogen atom mass. Note that to avoid numerical artifacts arising from the convolution over the 21-cm absorption line profile, one needs to make sure that $\delta v<b$. In this study, this was achieved by resampling the LoS of IGM fields using linear interpolation (see \citealt{Soltinsky_2021} for more details). Given the expected large amount of the \Lya~photons in the universe at $z=6$, which provide the coupling between the spin temperature, $T_{\rm S}$, and $T_{\rm K}$, there is little difference between these two temperatures \citep{Soltinsky_2021}, and hence $T_{\rm S}=T_{\rm K}$ is assumed. The normalized 21-cm forest flux spectrum is defined as $F_{21}=\rm e^{-\tau_{21}}$. To construct longer 21-cm forest flux spectra, four randomly chosen original spectra were spliced together extending the spectra to $200\,\rm cMpc$ or equivalently to $22.1\,\rm MHz$ at $z=6$.

Following the \citet{Soltinsky_2025}, we incorporate an intrinsic radio spectrum of the background source and instrumental features including the spectral resolution of $\Delta\nu=8\,\rm kHz$ and thermal noise of the telescope. The intrinsic spectrum was modeled as a single power-law, i.e. $S=S_{147}\left(\nu/147\,\mathrm{MHz}\right)^{\alpha_{\rm R}}$, where $\alpha_{\rm R}$ is the radio spectrum index and $S_{147}$ is the intrinsic flux density at $147\,\rm MHz$. 
We assume the values of $S_{147}=64.2\,\rm mJy$ and $\alpha_{\rm R}=-0.44$ which correspond to the $z=6.1$ blazar PSO J0309+27 \citep{Belladitta_2020}.
The thermal noise of the telescope is modeled as Gaussian white noise with the rms computed from the radiometer equation \citep[cf.][]{Datta_2007,Ciardi_2013}
\begin{equation}
    \sigma_{\rm N} = \frac{T_{\rm sys}}{A_{\rm eff}}\frac{\sqrt{2}k_B}{\sqrt{\Delta\nu t_{\rm int}}},
    \label{eq:noise}
\end{equation}
where $A_{\rm eff}/T_{\rm sys}$ is the sensitivity of the telescope assuming the whole array (i.e. 30 dishes and 512 stations in the case of the uGMRT and SKA1-low, respectively) given in \citet{Braun_2019} and $t_{\rm int}$ is the integration time. An example of transmission spectra is in Fig.~\ref{fig:denoising-gmrt50h}.

\subsection{The 1D transmission power spectrum}\label{sec:1dpower_spectrum}
In this study we consider a 1D power spectrum of the 21-cm forest normalized flux, $P_{21}$, as one of the observables. Assuming the estimator $\delta_{\rm F}=F_{21}-1=\rm e^{-\tau_{21}}-1$, this observable is defined as
\begin{equation}
    P_{21}\left(k\right)\delta_{\rm D}\left(k-k'\right)=\left\langle\tilde{\delta}_{\rm F}\left(k\right)\tilde{\delta}_{\rm F}^*\left(k'\right)\right\rangle \label{eq:PS_def}
\end{equation}
\noindent
and using discrete Fourier transforms of our estimator, $\tilde{\delta}_{\rm F}$, it is estimated for a 21-cm forest spectrum with the number of pixels equal to $n$ as \citep{Khan_2024}
\begin{equation}
    P_{21}\left(k_q\right)=\left(\frac{2\pi}{n\Delta\nu}\right)\langle\lvert\tilde{\delta}_{\rm F}\left(k_q\right)\rvert^2\rangle,\label{eq:PS_comp}
\end{equation}
\noindent
where $\delta_{\rm D}$ is the Dirac delta function, $k$ is the wavenumber and $k_q=2\pi q/n\Delta\nu$ for $q=0,1,...,n-1$.

Throughout this study, for analysis we use the binned dimensionless product $kP_{21}$, examples of which can be found in Fig.~\ref{fig:difficult-space-ps}. We have binned it over the range $10^{-0.5}\mathrm{MHz}^{-1} \leq k \leq 10^{2.5}\mathrm{MHz}^{-1}$ in the logarithmic $k$-bins of size 0.25 dex. 

\section{Likelihood-based Bayesian inference}\label{sec:MCMC_noisesubtracted}

Before introducing our machine learning approaches, we first revisit the Bayesian inference method based on the 1D power spectrum of the 21-cm forest signal, as described in \citet{Soltinsky_2025}. We refer to this as Method~A1. As an improvement, we also consider a variant in which the noise contribution is subtracted from the 21-cm forest power spectrum prior to inference; we refer to this as Method~A2. Both methods are illustrated in the flowchart in Fig.~\ref{fig:five-inference-pipelines}.

\begin{figure*}
    \centering
\includegraphics[width=0.82\linewidth]{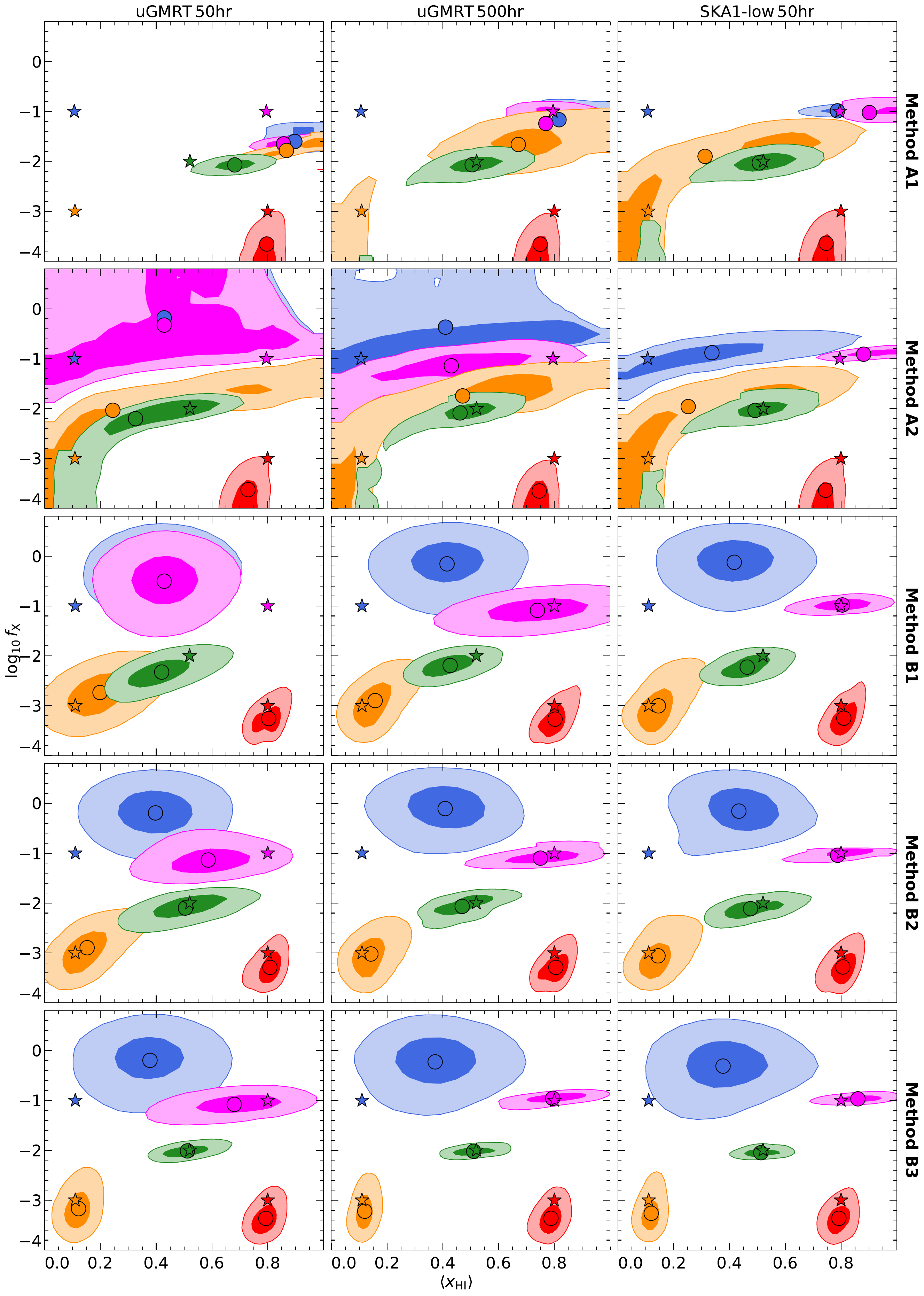}
    \caption{Posterior distributions for the two model parameters, $\mathrm{log}_{10}f_{\rm X}$ and $\langle x_{\rm HI}\rangle$, inferred from mock 21-cm forest spectra with instrumental noise at $z=6$. Contours indicate the $1\sigma$ and $2\sigma$ credible intervals in the two-dimensional parameter space. Stars mark the true parameter values, while circles show the median inferred values for each test case. Columns correspond to increasing telescope sensitivity (from left to right, uGMRT 50~hr, uGMRT 500~hr, and SKA1-low 50~hr), while rows show progressively advanced inference methods (from top to bottom, methods A1, A2, B1, B2, and B3).}
    \label{fig:posterior-plots-all}
\end{figure*}

 \begin{table*}
     \centering
     \renewcommand{\arraystretch}{1.2}
     \begin{tabular}{cllccc}
 \toprule
 \multicolumn{3}{c}{} & \multicolumn{3}{c}{\textbf{Effectiveness score} $E$ (higher means better)} \\
 \cmidrule(r){4-6}
 &
 \textbf{Method used} & \textbf{Feature set} & \textbf{uGMRT 50~hr} & \textbf{uGMRT 500~hr} & \textbf{SKA1-low 50~hr} \\
\midrule
A1 & Bayesian inference  & Power spectrum of noisy spectrum       & 0.00 & 0.00 & 0.00 \\
A2 & Bayesian inference  & Power spectrum with noise subtraction & 1.50 & 2.67 & 4.69 \\
\midrule
B1 & ML Regression       & Power spectrum of noisy spectrum      & 0.54 & 2.15 & 3.66 \\
B2 & ML Regression       & Power spectrum of U-Net denoised flux & 1.98 & 3.06 & 4.28 \\
B3 & ML Regression & Latent features from the U-Net & 4.06 & 5.73 & 5.81 \\
\bottomrule
    \end{tabular}
\caption{Effectiveness score, $E$, defined by equation~(\ref{eq:merit}), for our five inference methods and three observation scenarios. Higher values of $E$ indicate better performance, as discussed in Section~\ref{sec:method_a1}. The methods are ordered by increasing effectiveness, with Method~A1 ranking lowest due to its limited ability to recover the true parameters with noisy observations. The best performance is achieved by Method~B3, which uses latent features from the U-Net instead of the transmission power spectrum.}
\label{tab:inference_comparison}
\end{table*}

\subsection{Method~A1: Inference without noise subtraction}
\label{sec:method_a1}

This method is presented in \citet{Soltinsky_2025}.  The observable is defined as the average power spectrum $\langle P_{21}\rangle_{10}$ computed as the mean over ten randomly selected sightlines at $z = 6$.  The goal is to infer the posterior distribution of our model parameters $\bm{\theta}=\{\mathrm{log}_{10}f_{\rm X},\langle x_{\rm HI}\rangle\}$.  In this likelihood-based Bayesian approach, which we term Method~A1, we perform this inference by writing a Gaussian likelihood
\begin{equation}
    \mathcal{L}(P_{21}\lvert\bm{\theta})=\frac{1}{\sqrt{\mathrm{det\,\textbf{C}}}}\exp\left(-\frac{1}{2}\mathrm{\textbf{d}^T\textbf{C}^{-1}\textbf{d}}\right),\label{eq:likelihood_covar}
\end{equation}
where the quantity $\textbf{d}$ is the difference between the chosen mock observation $\langle P_{21}^{\rm mock}\rangle_{10}$ and the simulated power spectrum $\langle P_{21}^{\rm sim}\rangle_{10}(\bm{\theta})$ at the parameter values $\bm{\theta}$, i.e. 
\begin{equation}
\textbf{d}=\langle P_{21}^{\rm mock}\rangle_{10}-\langle P_{21}^{\rm sim}\rangle_{10}(\bm{\theta}). 
\end{equation}
In this method we assume that mock power spectra are computed from noisy 21-cm forest flux spectra, i.e. $\langle P_{21}^{\rm mock}\rangle_{10}=\langle P_{21}^{\rm S+N}\rangle_{10}$, where S and N correspond to the signal and noise contributions, respectively.
The covariance matrix $\textbf{C}$ is constructed from an ensemble of $10^4$ realizations of the $\langle P_{21}^{\rm S+N}\rangle_{10}$. Hence, both the sample variance and instrumental noise are incorporated in the $\textbf{C}$. The posterior distribution is then given by 
\begin{equation}
    P(\bm{\theta}\lvert P_{21}^{\rm mock})\propto\mathcal{L}(P_{21}^{\rm mock}\lvert\bm{\theta})\mathcal{P}(\bm{\theta}),\label{eq:bayes_theorem}
\end{equation}
where $\mathcal{P}(\bm{\theta})$ is the prior distribution of the parameters, which is assumed to be uniform in this case. The posterior distribution is sampled using the \texttt{emcee} code \citep{Foreman-Mackey_2013_emcee} via the MCMC method \citep{Goodman_2010}.

We show the results of this approach in the top row panels in Fig.~\ref{fig:posterior-plots-all}. The posterior distributions are shown as shaded regions, with darker shades corresponding to the highest posterior density $1\sigma$ ($39.3\%)$ credibility intervals and lighter shades marking the $2\sigma$ ($86.4\%)$ credibility intervals\footnote{For the discussion on the confidence intervals in a 2D distribution we refer the reader to \hyperlink{https://corner.readthedocs.io/en/latest/pages/sigmas/}{https://corner.readthedocs.io/en/latest/pages/sigmas/}}. The circles indicate the best fit parameter values ($\mathrm{log}_{10}f_{\rm X}^{\rm inf}$, $\langle x_{\rm HI}\rangle^{\rm inf}$) resulting from this MCMC procedure, while the stars represent the (true) values of parameters assumed for the mock observation ($\mathrm{log}_{10}f_{\rm X}^{\rm true}$, $\langle x_{\rm HI}\rangle^{\rm true}$). We perform the analysis on five test models spread out in the parameter space.  Going from left to right, the top panels show the results for the mock observations by the uGMRT over $50\,\rm hr$, uGMRT over $500\,\rm hr$ and SKA1-low over $50\,\rm hr$, i.e. increasing sensitivity. Note that throughout the paper $t_{\rm int}$ are given per source. The posterior distributions are not able to encompass the true parameter values for any of the test models, especially in the case of uGMRT over $50\,\rm hr$.  In the case of the parameter point with $[\langle x_{\rm HI}\rangle^{\rm true}, \mathrm{log}_{10}f_{\rm X}^{\rm true}]=[0.11, -1]$ (blue colour), the posterior distribution fails to encompass the true parameter values even for the SKA1-low over $50\,\rm hr$ mock observation. 

We quantify the performance of our inference methodology through an effectiveness score, $E$. This is defined based on the probability assigned to the true values of our model parameters in the posterior distributions.  If the posteriors peak on the true values and are narrow, the effectiveness score is higher.  In order to compute this, we fit a Gaussian kernel density estimator (KDE) model to our posterior distributions around known test points. Specifically, for each of the five test points $(x_i^\ast,y_i^\ast)$ ($i=1,\dots,5$), we interpret the corresponding posterior sample as independent draws from an unknown bivariate density $f_i(x,y)$. We estimate this density using an isotropic two-dimensional Gaussian KDE at position $\mathbf{r} = (x, y)$,
\begin{equation}
\hat f_i(\mathbf r)\;=\;
\frac{1}{N h_x h_y}\sum_{j=1}^{N}
\exp\;\left(-\frac12
\left[\frac{(x - x_{i,j})^2}{h_x^{2}} 
     + \frac{(y - y_{i,j})^2}{h_y^{2}}\right]
\right),
\end{equation}
where $\hat f_i(\mathbf r)$ denotes the estimated probability density at $\mathbf{r}$ for the $i$th test case, $N$ is the total number of samples, $(x_{i,j}, y_{i,j})$ are the coordinates of the $j$th sample, and $h_x$, $h_y$ are the kernel bandwidths that control the smoothing along the $x$ and $y$ directions. This expression effectively places a two-dimensional Gaussian centered at each sample point, and summing over all $N$ samples yields a smooth estimate of the underlying bivariate probability density.
The probability that a future sample from $f_k$ lies within a tile of width $\Delta x=0.05$ and height $\Delta y=0.2$\footnote{These values roughly correspond to the separation of the models in Figure~\ref{fig:training-test-points}. We arbitrarily choose these values so that we can work with dimensionless probabilities.} centred on $(x_i^\ast,y_i^\ast)$ is then approximated by
\begin{equation}
p_i \;=\; \hat f_i(x_i^\ast,y_i^\ast)\,\Delta x\,\Delta y.
\end{equation}
We then compute the effectiveness score from these probabilities by computing the geometric mean across the five test points, and scaling it by a factor of $100$ for convenience:
\begin{equation}
E
= 100 \times 
\left(\prod_{i=1}^{5} p_i\right)^{1/5}.
\label{eq:merit}
\end{equation} 
This metric serves as our principal figure of merit for comparing the inference methods examined in this study. A higher value of $E$ indicates a more effective and reliable method. Table~\ref{tab:inference_comparison} summarizes the $E$ scores achieved by each inference approach across the different telescope configurations. In general, a higher $E$ signifies greater success in recovering the true parameter values. As expected, improvements in observational sensitivity lead to increased $E$ scores: moving from left to right across any row in Table~\ref{tab:inference_comparison}, corresponding to uGMRT observations over $50\,\rm hr$, uGMRT over $500\,\rm hr$, and SKA1-low over $50\,\rm hr$, we observe a consistent enhancement in $E$. An exception is Method~A1, the initial approach, for which $E$ remains zero up to a reasonable precision under all configurations, reflecting limited ability to recover the target parameters with noisy observations. This is also evident in Fig.~\ref{fig:posterior-plots-all}, where the posterior contours in the top row fail to encompass the true parameter values for most test models.

\subsection{Method~A2: Inference with noise subtraction}

In a spectrum of a radio-loud quasar, the flux measurement on the red side of the rest-frame 21-cm line can be considered as a measurement of the telescope noise.  This opens up the prospect of noise mitigation using the noise spectrum to improve the inference of the 21-cm forest signal. In practice, alternative approaches to noise measurement can also be available, such as measurements of a radio source at a lower redshift (i.e. post-reionization redshift at which no 21-cm forest absorption is expected) at the frequencies of interest.  Furthermore, instead of measuring the telescope noise, one can forward-model it and its corresponding power spectrum, and subtract this from the observationally measured noisy signal. In this section, we introduce a new method, Method~A2, which incorporates noise subtraction into the Bayesian inference process discussed above in Section~\ref{sec:method_a1}.

In this case, the inference procedure is broadly along the same lines as Method~A1.  The priors are uniform, and the likelihood is defined in the same way as in equation~(\ref{eq:likelihood_covar}). The $\textbf{C}$ includes both the sample and noise variance like in Method~A1. However, the quantity \textbf{d} is now defined as the difference between the noise-subtracted mock observation $\langle P_{21}^{\rm N_{\rm sub}}\rangle_{10}$ and the simulated power spectrum $\langle P_{21}^{\rm sim}\rangle_{10}(\bm{\theta})$ at the parameter values $\bm{\theta}$, i.e.
\begin{equation}
\textbf{d}=\langle P_{21}^{\rm N_{\rm sub}}\rangle_{10}-\langle P_{21}^{\rm sim}\rangle_{10}(\bm{\theta}).
\end{equation}
Here the noise-subtracted mock power spectrum is defined as
\begin{equation}\label{eq:noise_sub_PS}
\langle P_{21}^{\rm N_{\rm sub}}\rangle_{10}=\langle P_{21}^{\rm S+N}\rangle_{10}-\langle P_{21}^{\rm N}\rangle,
\end{equation}
where, $\langle P_{21}^{\rm N}\rangle$ is the mean 1D power spectrum computed over 1000 realizations of the noise-only spectra. The noise-only spectra are computed by taking the intrinsic flux spectrum of the radio source and adding Gaussian noise with the rms computed from the radiometer equation (equation~\ref{eq:noise}). Note that in equation~(\ref{eq:noise_sub_PS}) the same $\langle P_{21}^{\rm N}\rangle$ is subtracted when varying ($\mathrm{log}_{10}f_{\rm X}^{\rm true}$, $\langle x_{\rm HI}\rangle^{\rm true}$) as long as the same observational setup is assumed.

The results of Method~A2 are shown in the second row of panels in Fig.~\ref{fig:posterior-plots-all}. As for Method~A1, the posterior distributions are shown as shaded regions. The circles indicate the best fit parameter values, while the stars represent the (true) values of parameters. We perform the analysis on five test models spread out in the parameter space.  The three panels show the results for the mock observations by the uGMRT over $50\,\rm hr$, uGMRT over $500\,\rm hr$ and SKA1-low over $50\,\rm hr$, i.e. increasing sensitivity.  Compared to the Method~A1 (top panels), we see that all of the true parameter values (stars) are within the posterior distributions (shaded regions). This is a significant improvement in comparison to the approach used in \citep{Soltinsky_2025}.  This improvement reflects in the metric $E$ (Table~\ref{tab:inference_comparison}), which has now increased to 1.50 for the uGMRT over $50\,\rm hr$ observation, 2.67 for the uGMRT over $500\,\rm hr$ observation and 4.69 for the SKA1-low over $50\,\rm hr$ observation.
The reason for the significantly superior performance of Method~A2 relative to Method~A1 is discussed in Appendix~\ref{sec:likelihood_test}. In brief, noise is added in the transmission, whereas our observable is the mean power spectrum. This mismatch causes the model to drift away from the mean in the noisy case, biasing the inference. As a result, the assumed Gaussian likelihood is no longer a good description of the true likelihood. When we subtract the noise from the data, this bias is largely corrected, and the inference improves accordingly. Specifically, the noise-subtracted power spectrum $\langle P_{21}^{\rm N_{sub}}\rangle_{10}$ is much better described by the likelihood in equation~(\ref{eq:likelihood_covar}) than the noisy power spectrum $\langle P_{21}^{\rm S+N}\rangle_{10}$. However, even the noise-subtracted 21-cm forest power spectrum retains some non-Gaussianity, so a Gaussian likelihood is not an ideal choice \citep{Wolfson_2023}. Therefore, in the following sections, we explore machine learning methods that do not rely on any explicit likelihood function.

\begin{figure*}
    \centering
    \includegraphics[width=0.9\linewidth]{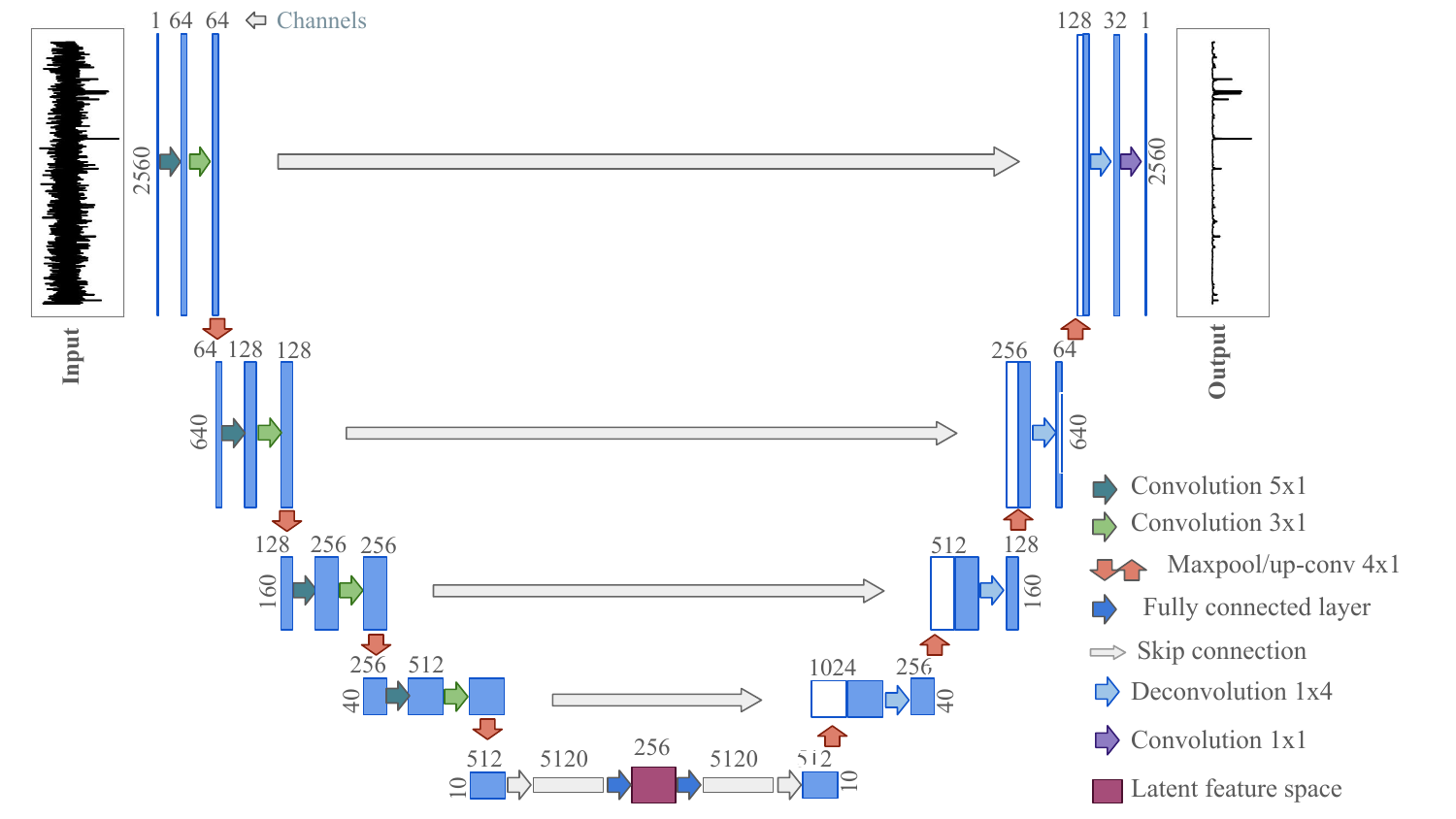}
    \caption{Schematic of the 1D U-Net architecture used for denoising 21-cm forest spectra. The encoder (left part of the `U') compresses the noisy input flux via successive convolution and max-pooling layers, while the decoder (right part of the `U') reconstructs the denoised output using upsampling and convolution. Skip connections preserve fine-scale spectral features by linking encoder and decoder layers at matching resolutions. The latent feature vector (bottom centre) is used for regression in Method~B3, while the denoised flux output is used for power spectrum computation in Method~B2.}
    \label{fig:u-net-architecture}
\end{figure*}

\section{Likelihood-free inference using machine learning}
\label{sec:Inference_ML}

We now turn to a likelihood-free inference approach using ML methods. Our goal, as before, is to infer the two parameters $\langle x_{\rm HI} \rangle$ and $\mathrm{log}_{10}f_{\rm X}$ from the noisy 21-cm forest flux spectra. We explore three different ML-based methods, which are illustrated in Fig.~\ref{fig:five-inference-pipelines} as Methods B1, B2 and B3. The first two methods use the 1D power spectrum of the noisy and denoised flux spectra, respectively, while the third method bypasses the power spectrum statistic entirely by using latent features extracted from a U-Net architecture.

\subsection{Method~B1: Inference using the 1D transmission power spectrum with XGBoost regression}
\label{sec:ml-inf-noisy}

As our first approach, we simply replace the MCMC inference method with a machine learning regression model, specifically XGBoost \citep{Chen2016}, to predict the two target parameters from the 1D power spectrum of the noisy flux. This method is illustrated as Method~B1 in Fig.~\ref{fig:five-inference-pipelines}.

Extreme Gradient Boosting (XGBoost) is one of the several `labeller' algorithms to emerge from the machine learning literature.  Such algorithms learn from examples, and can both classify (stick labels on) and regress (estimate numeric values) on data. XGBoost \citep{Chen2016} is a scalable and regularized implementation of gradient-boosted decision trees \citep{Friedman01} that balances model interpretability and predictive performance. Building on the gradient-boosting framework of \citet{Friedman01}, \citet{Chen2016} introduced XGBoost with three enhancements: (i) use of a second-order Taylor expansion of the loss function to improve split evaluation, (ii) inclusion of $L_1$ and $L_2$ regularization to control model complexity without relying on pruning, and (iii) a sparsity-aware, block-compressed data format that supports efficient parallel and out-of-core computation on both CPUs and GPUs. The method constructs an ensemble of shallow regression trees, each trained to minimize the residuals of the previous model. Regularization techniques such as shrinkage, column subsampling, and a tunable learning rate allow fine control over model complexity and generalization. XGBoost has been applied in astronomy for star-galaxy classification \citep{2019ChA&A..43..539L, 2025MNRAS.537..876A, 2024MNRAS.527.4677Z}, photometric redshift estimation \citep{2022A&A...666A..87C}, halo-galaxy connection modelling \citep{2021MNRAS.507.1468M, 2019MNRAS.490.2367C}, and transient classification \citep{2022A&A...665A..99M}. While convolutional neural networks are well-suited to image-based data, XGBoost is particularly effective when the predictive information resides in structured, tabular feature vectors.\footnote{The official GPU-enabled implementation of XGBoost is maintained at \url{https://github.com/dmlc/xgboost}.}

In our case, for Method~B1, the training set for our XGBoost model consists of 529 models shown in Fig.~\ref{fig:training-test-points}, each corresponding to a unique pair of $\langle x_{\rm HI} \rangle$ and $\mathrm{log}_{10}f_{\rm X}$. For each model, we generate 1000 noisy 21-cm forest sightline transmission flux spectra ($F^{\rm S+N}_{21}$) as described in Section~\ref{sec:21cm_forest}. We select 200 spectra from each model as training samples. That gives us a total of $529 \times 200 = 105,800$ LoS samples. We then compute the corresponding 1D power spectra for these noisy flux spectra using the method described in Section~\ref{sec:1dpower_spectrum}, each 1D power spectrum array containing the value of $kP_{21}$ in 16 logarithmic $k$-bins, as the feature vectors. To increase robustness of the inference, we take the average of 10 randomly selected feature vectors as a training sample. This results in a total of $10,580$ training samples. These feature vectors are then used to train the XGBoost regressor, with the known values of the two target parameters provided as labels. 

We use the default loss function in XGBoost, which is the Mean Squared Error (MSE):
\begin{equation}
\mathrm{MSE} = \frac{1}{N_{\rm params}} \sum_{i=1}^{N_{\rm params}} \left( y^{\rm pred}_i - y^{\rm true}_i \right)^2,
\end{equation}
where $N_{\rm params}=2$ is the number of parameters to be inferred, $y^{\rm pred}_i$ is the predicted value of the $i$-th parameter, and $y^{\rm true}_i$ is the true value of the $i$-th parameter.
The default \texttt{gbtree} booster is used, which builds an ensemble of decision trees to fit the training data. The model is trained to minimize the loss function at each step by adding and updating decision trees. We fixed the random seed of XGBoost (\texttt{random\_state}), which allows us to reproduce the results when run multiple times with the same data. Rest of the hyperparameters are left unspecified. The default values for the most important parameters are: $\texttt{n\_estimators}=100$, $\texttt{max\_depth}=6$, and $\texttt{learning\_rate}=0.3$. We used XGBoost version 2.1.1 with Python 3.12.2 for the implementation. Training was performed on an Apple M2 CPU (MacBook Air), and took less than a minute. The final trained model is archived in the project GitHub repository and is available for download.

The trained XGBoost model is then tested on five representative parameter combinations spanning the parameter space (test points, as shown in Fig.~\ref{fig:training-test-points}). 
Here is an important note on how we achieved a clean separation of training and test data. In our simulation, we have mocked 1,000 LoS in each model, which represent identical locations in space but with different physical parameters in different models. We have isolated our training from test data on both dimensions - LoS and physical parameters.
For training, we included only 200 LoS from each of the 529 training models. For testing, we used the set of 800 LoS that are excluded while training, from each of the 5 testing models. That gives us a total of $800 \times 5 = 4,000$ LoS samples for testing. 

To derive a posterior distribution over the physical parameters from the trained XGBoost regressor, we construct a synthetic test dataset. For each of the five representative test models shown in Fig.~\ref{fig:training-test-points}, we map the 800 testing LoS spectra to 1D power spectra. Using these 800 power spectra, we perform a bootstrap procedure to construct $10^4$ input feature vectors by resampling groups of 10 random power spectra. Each of these feature vectors is then passed through the trained XGBoost model to obtain corresponding predictions for the two physical parameters of interest.
The ensemble of $10^4$ predictions per test model is treated as a sample from an approximate posterior distribution $p(\bm{\theta}|\mathbf{d}_{\text{test}})$, where $\bm{\theta}$ denotes the parameters and $\mathbf{d}_{\text{test}}$ denotes the observational features derived from the flux spectra. From this empirical distribution, we compute the posterior mean and identify the regions enclosing 39.3\% and 86.4\% of the predicted samples, corresponding to the $1\sigma$ and $2\sigma$ credible intervals, respectively. These are then visualized as two-dimensional confidence contours in the parameter space. The medians of predicted parameter values are considered as the best-fit.  All the ML-based inference methods follow this procedure for creating posterior distributions.

The third row of Fig.~\ref{fig:posterior-plots-all} plots the posterior distributions of this method and the third row of Table~\ref{tab:inference_comparison} shows the corresponding $E$ scores. Compared to Method~A2, this method shows drastically different behaviour in two parts of the parameter space. It shows remarkable improvement for the  test models where the signal is strong, at lower values of $\mathrm{log}_{10}f_{\rm X}^{\rm true}$. The contours for these points are narrow and they encompass the true values. The improvement is especially notable for the point $[\langle x_\mathrm{HI}\rangle , \mathrm{log}_{10}f_{\rm X}] = [0.11, -3.0]$ (orange marker). While this point is poorly inferred even with SKA1-low data under the Bayesian framework, it is well-recovered with ML inference. 
As we go from left to right, towards more sensitive observations, the performance improves significantly, $E$ increasing from 0.54 to 3.66. In case of uGMRT over $50\,\rm hr$, only three test points were encompassed by the posterior contours. In the middle and right panel, four of five test points are encompassed. 
This part of the result underscores the effectiveness of XGBoost in capturing complex relationships between the power spectra and the physical parameters, during regression. To gain further insight into how XGBoost utilizes input features to create the regression model, we refer the reader to Appendix~\ref{sec:feature-imp} which presents the relative importance of individual power spectrum bins as determined by XGBoost. 

However, the two test models at $\mathrm{log}_{10}f_{\rm X}^{\rm true} = -1$ perform poorly for uGMRT over $50\,\rm hr$. These two posterior contours overlap each other and do not encompass their true values. For the more sensitive telescopes, this issue persists for one test point. $E$-score calculation penalizes this behaviour severely, resulting in lower scores for this method compared to Method~A2 but better than Method~A1. 

In short, XGBoost works extremely well when the input features have a sharp contrast among different test models. However, it performs poorly when the input features have dominating noise, wiping out the contrast. Fig.~\ref{fig:difficult-space-ps} shows an example each of sharp contrast, medium contrast and no contrast, for power spectrum features used in Method~B1. More discussion on this swing in XGBoost performance in two different parameter ranges can be found in Section~\ref{sec:conclusion}. Our endeavour in the subsequent sections would be to extract features that provide a sharper contrast.

\subsection{Method~B2: Inference using U-Net denoising}
\label{sec:ml-inf-denoised}

It may seem that using the noise-subtracted power spectrum, as done in Method~A2, would improve the regression with XGBoost. However, when we tried this, it showed no improvement over Method~B1. This implies that the XGBoost is able to automatically detect the underlying average noise spectrum during training. So we consider an alternative denoising procedure. We first denoise the noisy flux spectra using a U-Net architecture, and then compute the 1D power spectrum of the denoised flux. The denoised power spectrum is then used to infer the parameters using XGBoost regression. This method is illustrated as Method~B2 in Fig.~\ref{fig:five-inference-pipelines}.

U-Net is a convolutional neural network architecture originally proposed by \citet{ronneberger2015unetconvolutionalnetworksbiomedical} for biomedical image segmentation. It enhances the classical encoder-decoder framework with skip connections that directly link each layer in the encoder to its corresponding layer in the decoder. This structure preserves spatial detail lost during downsampling, enabling the network to recover fine-scale features while retaining global contextual information. The encoder compresses the input through successive convolution and pooling layers, while the decoder reconstructs the output via transpose convolutions, aided by concatenated high-resolution features from the encoder. U-Net's capacity to retain small-scale structure makes it especially effective in astrophysical applications where faint, spatially localized signals must be separated from complex noise and backgrounds. It has been employed in tasks such as de-blending galaxy images \citep{2024PASA...41...35Z, 2020MNRAS.491.2481B}, 21-cm foreground subtraction \citep{2021JCAP...04..081M}, and denoising diffuse radio maps \citep{2024MNRAS.533.3194S}. In the context of 21-cm forest analyses, U-Net serves as a robust architecture for denoising sightline data and reconstructing coherent absorption features. Moreover, the compressed representation at the bottleneck—the so-called latent space—can be leveraged to extract physically meaningful patterns and enable downstream tasks such as parameter estimation or model classification. Open-source reference implementations of the U-Net architecture are available in both Keras and PyTorch.

Here is a brief note on how we chose the input dimension for our U-Net design. Our simulated spectrum (see Section~\ref{sec:simulation}) consists of 2762 frequency channels. This irregular input size poses practical challenges for the U-Net architecture, as each downsampling step reduces the input length by a factor of 4. When the input length is not divisible by powers of 4, repeated downsampling leads to edge cropping and misalignment, requiring non-trivial padding strategies during upsampling to restore the original resolution. These corrections increase architectural complexity and computational overhead. To circumvent these issues, we truncate the input and use only the first 2560 points of the spectrum, ensuring compatibility with the U-Net structure while preserving the vast majority of the spectral information. This truncation does not significantly affect the overall analysis, as the majority of the signal information is retained.

The U-Net architecture used in this study (Fig.~\ref{fig:u-net-architecture}) maps the noisy flux $F^{\rm S+N}_{21}$ to its noise-free counterpart $F^{\rm D}_{21}$, where the superscript `D' denotes `denoised'. The architecture follows a standard encoder-decoder design with skip connections to preserve spatial detail. The encoder consists of four one-dimensional convolutional layers, each using two kernels of sizes 5 and 3 respectively, stride 1, and ReLU activation, followed by max pooling for downsampling. Small kernel sizes were selected considering the need to extract fine grained details from the spectrum. After each convolution layer, we use batch normalization to provide training stability. The ReLU activation function introduces non-linearity, which is crucial for learning complex patterns in data. We also used dropout regularization of 20\% after each convolution layer, which prevents overfitting. Downsampling by a factor of 4 allows us to compress the 2560 length input vector to a compact representation. Through successive convolution and pooling operations, the input spectrum is compressed to length 10, while the number of feature channels increases from 1 to 512. This high-dimensional representation is then flattened and projected into a 256-dimensional latent feature space via a fully connected layer. These latent features are designed to encode the essential characteristics in the input flux spectrum.

The decoder mirrors the encoder structure, employing upsampling layers followed by convolutional operations to reconstruct the spectrum at the original input resolution from the latent features. We used transposed convolution layers for upsampling which have learnable parameters and are more adaptable compared to interpolation methods. Skip connections between encoder and decoder layers at matching resolutions ensure that fine-scale information is propagated during reconstruction. This allows the network to optimize the latent representation for the essential, high level features, and still be able to localize fine-grained features in the output spectrum.

\begin{figure}
    \centering
    \includegraphics[width=1\linewidth]{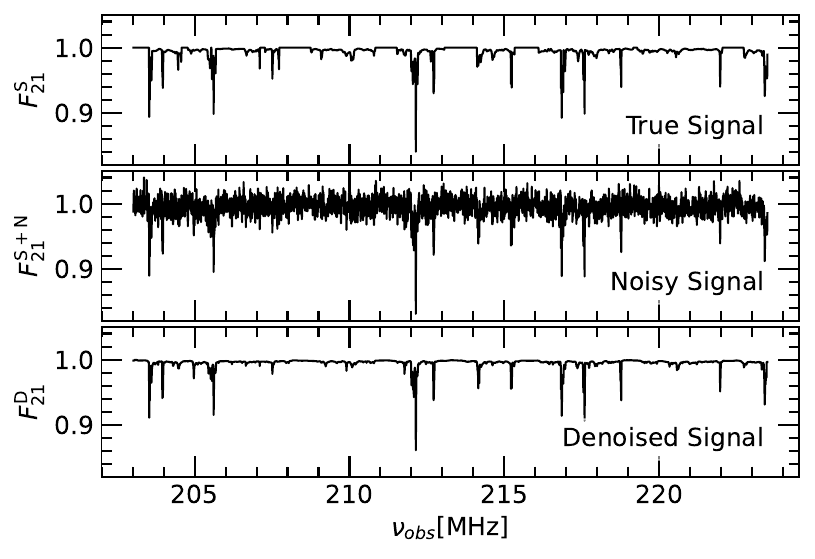}
    \caption{
    Example of U-Net denoising performance for a representative 21-cm forest sightline at $[\langle x_{\rm HI}\rangle, \mathrm{log}_{10}f_{\rm X}] = [0.80, -3]$. The top panel shows the true (noise-free) absorption signal ($F_{21}^{\rm S}$), the middle panel shows the noisy flux simulating a uGMRT $50\,\rm hr$ observation ($F_{21}^{\rm S+N}$), and the bottom panel shows denoised flux extracted by the U-Net ($F_{21}^{\rm D}$).
    The U-Net successfully recovers the underlying absorption features, substantially reducing the impact of instrumental noise.
    }
\label{fig:denoising-gmrt50h}
\end{figure}

Rather than reconstructing the noisy input spectrum, we train the model in a denoising autoencoder setting, to predict the true (i.e., noise-free) flux signal from the noise-contaminated input. This is achieved using the Mean Relative Squared Error (MRSE) loss function,
\begin{equation}
\mathrm{MRSE} = \frac{1}{N_{\rm freq}} \sum_{i=1}^{N_{\rm freq}} \frac{(F^{\rm pred}_i - F^{\rm true}_i)^2}{F^{\rm true}_i},
\label{eq:msre}
\end{equation}
where $F^{\rm pred}_i$ is the predicted and $F^{\rm true}_i$ the true (simulated) transmission in the $i$-th frequency bin, and $N_{\rm freq}$ is the number of frequency points in the spectrum (2560). This loss formulation is preferable to plain MSE for spectra with significant fluctuation due to noise, as in the case of the 21-cm forest.

We train the U-Net model with $200$ LoS per training model, resulting in $529 \times 200$ training examples. The network is optimized using the Adam optimizer \citep{kingma2014adam}, a first-order gradient-based method that adapts the learning rate for each parameter based on estimates of first and second moments of the gradients. Adam is widely regarded as a reliable and efficient optimizer, often requiring minimal hyperparameter tuning. In our experiments, we set the learning rate to 0.0001 and the batch size to 32 based on empirical performance, while all other hyperparameters were kept at their default values. We evaluated the test set using the MSRE loss function (equation~\ref{eq:msre}) after every 5 epochs. Although the training loss continued to decrease, the loss on the test set reached a minimum of 34.09 after 15 epochs and then stopped decreasing, indicating over-fitting. Hence the training was terminated at 15 epochs. 
The U-Net model we have designed is relatively small and can be trained in a reasonable time on a CPU with a server class machine. The training was performed on a Linux server with 96GB of available memory. It took 1 hour, 36 minutes of wall time to execute on the CPU. It is worth noting that the entire ML pipeline we have developed - U-Net + XGBoost - does not put a heavy demand on computational resources. We have used Python version 3.12.2 and Pytorch version 2.5.1 in this work. The final trained model is archived in the project GitHub repository and is available for download.

\begin{figure}
    \centering
    \includegraphics[width=1\linewidth]{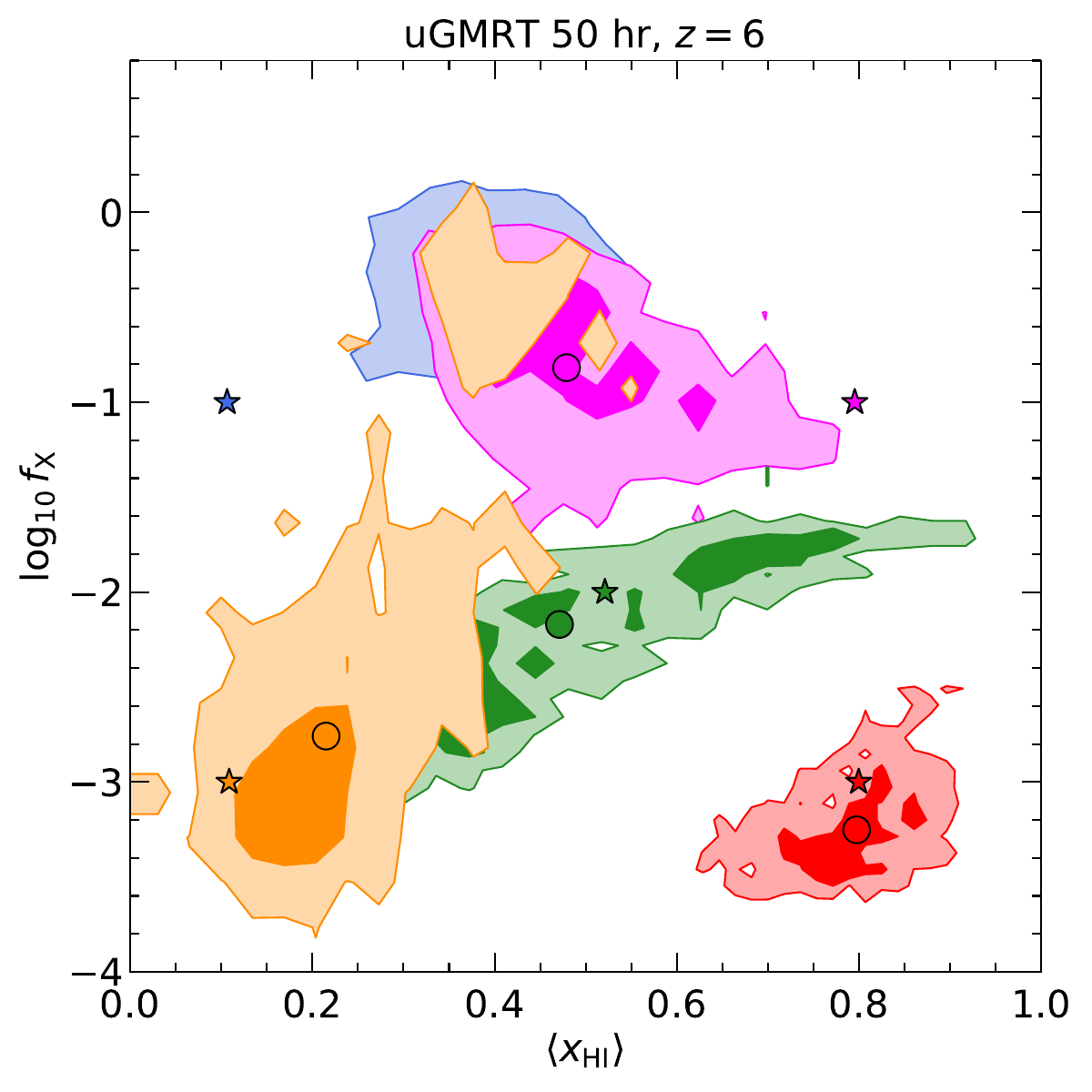}
    \caption{Posterior distributions for inference from a single sightline using latent features (Method~B3) for a 50~hr observation on uGMRT. This is the worst case observational scenario considered in this study. The contours are broader due to increased sample variance, but meaningful constraints on the parameters are still achieved.}
    \label{fig:inference-unet-single-los-gmrt50h}
\end{figure}

Fig.~\ref{fig:denoising-gmrt50h} demonstrates the performance of our ML-based denoising pipeline on a representative simulated uGMRT observation. Visual inspection indicates that the U-Net architecture recovers a significant fraction of the underlying 21-cm absorption features across a range of signal-to-noise scenarios. 
The effectiveness of our reconstruction is quantitatively confirmed by substantial reductions in MSRE (equation~\ref{eq:msre}). 
For instance, at a representative parameter combination $\langle x_{\rm HI}\rangle=0.52$, $\mathrm{log}_{10}f_{\rm X}=-2.0$, our denoised spectra achieve a MSRE of 0.02, a dramatic improvement over the initial noisy spectra (MSRE = 0.37). However, some limitations of U-Net denoising are also noted. It fails to fully reconstruct the subtle spectral features, e.g. features between 207 and 208~MHz in Fig.~\ref{fig:denoising-gmrt50h}. This leads to an over-subtraction in the power spectrum and introduces more uncertainty in it. More discussion on this can be found in Section~\ref{sec:conclusion}.

\begin{figure*}
    \centering
    \includegraphics[width=0.7\linewidth]{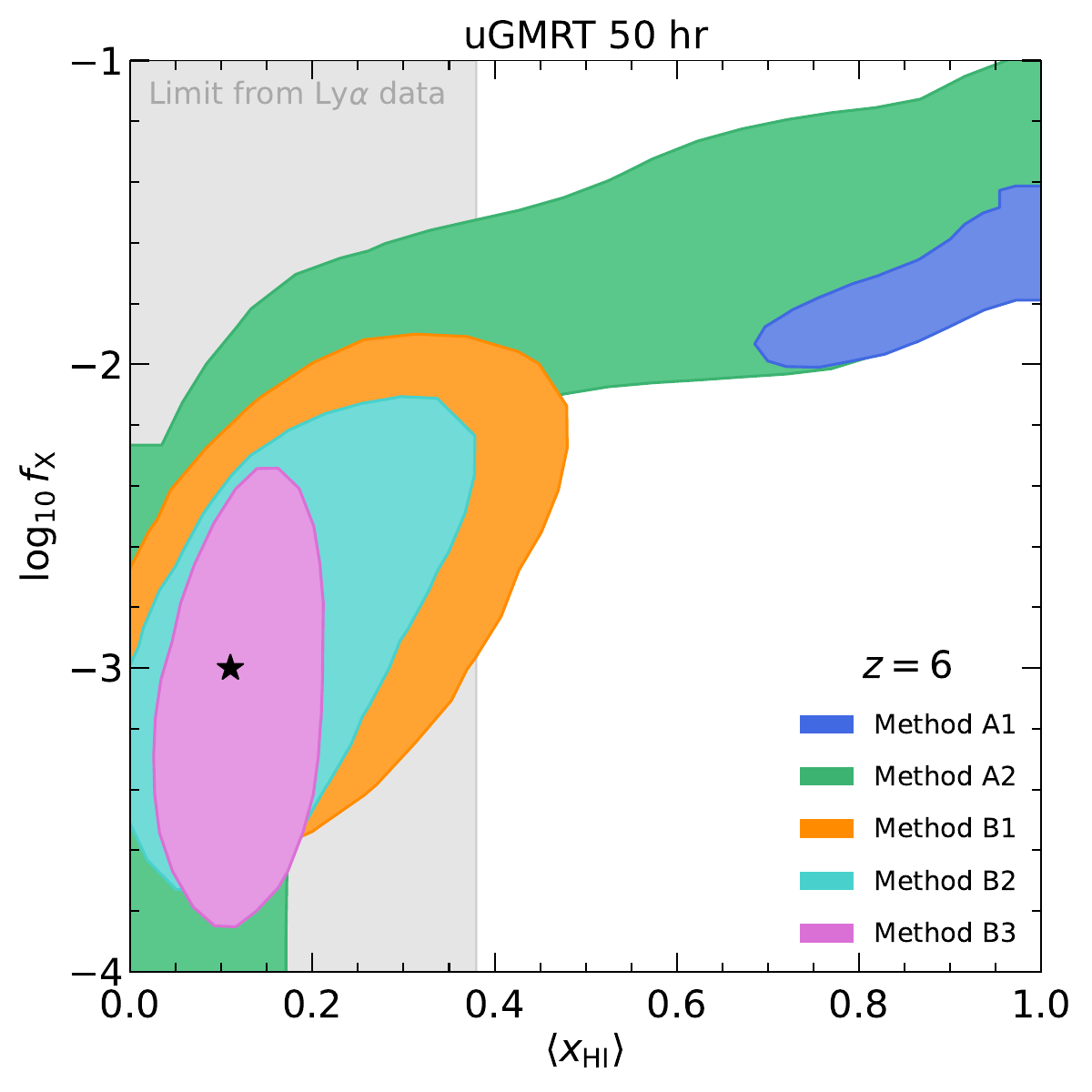}
    \caption{
    Comparison of all inference methods for a representative test point $[\langle x_{\rm HI}\rangle, \mathrm{log}_{10}f_{\rm X}] = [0.11, -3]$ using a mock uGMRT 50~hr observation. The black star marks the true parameter values, while coloured circles and contours show the mean estimates and corresponding $2\sigma$ credible intervals for each method. A clear progression is evident from Method~A1 to Method~B3, with posterior contours becoming progressively narrower and more accurately centred on the true values. The grey shaded region represents the limits of $\langle x_{\rm HI}\rangle\lesssim0.38$ from the Ly$\alpha$ forest observations at $z\sim 6$ (\citealt{2015MNRAS.447..499M}, \citealt{2022ApJ...932...76Z}, \citealt{2023MNRAS.525.4093G}, \citealt{2024ApJ...969..162D}, \citealt{2024MNRAS.530.3208G}).}
    \label{fig:inference-multi-method}
\end{figure*}

Once we have denoised the 21-cm forest transmission spectra using the U-Net, the inference process is the same as in Method~B1. We compute the 1D power spectrum of the denoised flux $F^{\rm D}_{21}$ using the method described in Section~\ref{sec:1dpower_spectrum}. 
The resulting 1D power spectrum $P^{\rm D}_{21}$ is then used as input to the XGBoost regressor. The trained XGBoost model predicts the two target parameters $\langle x_{\rm HI} \rangle$ and $\mathrm{log}_{10}f_{\rm X}$ for each of the five test points, and we perform a bootstrap procedure to construct an approximate posterior distribution over these parameters.  

Our results for Method~B2 are shown in the fourth row of Fig.~\ref{fig:posterior-plots-all} and  Table~\ref{tab:inference_comparison}. 
We see that Method~B2 achieves an $E$-score higher than Method~B1 for all three telescope configurations, and higher than Method~A2 for two out of three telescope configurations. Visually, we can see the point $[\langle x_{\rm HI}\rangle^{\rm true}, \mathrm{log}_{10}f_{\rm X}^{\rm true}] = [0.80, -1.0]$ (pink) is now captured by the posterior contours even for uGMRT over $50\,\rm hr$. The sizes of the contours are smaller than Method~B1 in general. However, we observe that the point $[\langle x_{\rm HI}\rangle^{\rm true}, \mathrm{log}_{10}f_{\rm X}^{\rm true}] = [0.11, -1.0]$ (blue) continues to elude the inference. As discussed above, the over-subtraction in power spectrum in some regions of the parameter space deteriorates the inference for these configurations.

\subsection{Method~B3: Inference using latent-space U-Net encoding}\label{sec:ml-inf-latent}

We now turn our attention to our most novel inference procedure, which, as we will see below, also turns out to be the best performing. In this method, termed Method~B3, we do not use the transmission power spectrum at all. Instead, we leverage the trained U-Net model described in Method~B2, not as a denoising tool but as a feature extractor. Specifically, we utilize the encoder (contracting path) of the U-Net to transform the input flux spectrum into a latent feature vector. During training, the encoder learns to extract high-level features from the input, capturing complex, non-Gaussian structures and hierarchical patterns that are often missed by traditional summary statistics like the power spectrum. These latent features serve as a highly informative input for regression, as they encode subtle correlations and non-linear relationships present in the data. Conceptually, the latent features are analogous to, say, the principal components in Principal Component Analysis(PCA), but the transformation is highly non-linear and data-driven, performed by the convolutional neural network. This makes the latent representation attractive for downstream inference tasks.

Similar to Methods~B1 and B2, the training set for our XGBoost model consists of 200 LoS from each of the 529 parameter combinations shown in Fig.~\ref{fig:training-test-points}. The key difference is that, for Method~B3, the feature vector for each training sample is the 256-dimensional latent representation extracted from the U-Net encoder, rather than the 16-dimensional power spectrum. For each model, we construct 200 latent feature arrays as training samples. As in previous methods, we further increase robustness by averaging the latent features over 10 randomly selected LoS, resulting in a total of $529 \times 20$ training samples, each of length 256.

Bottom rows of Fig.~\ref{fig:posterior-plots-all} and Table~\ref{tab:inference_comparison} present the results for this method. The posterior distributions are notably narrower and consistently centred on the true parameter values, demonstrating a clear improvement over all previous inference approaches across the three telescope configurations. The effectiveness score $E$ is substantially higher than for the other methods, ranging from 4.06 to 5.81. Even for the most challenging test case, $[\langle x_{\rm HI}\rangle^{\rm true}, \mathrm{log}_{10}f_{\rm X}^{\rm true}] = [0.11, -1]$, the inference accuracy is markedly improved. Also, interestingly, the performance gap between the least sensitive configuration (50~hr with uGMRT) and the more sensitive setups is significantly reduced.

To test this method in an extreme case, we also applied Method~B3 to a mock observation consisting of a single sightline observation from uGMRT over $50\,\rm hr$. The result is shown in Fig.~\ref{fig:inference-unet-single-los-gmrt50h}. Although the posterior contours are broader due to increased sample variance, the method still yields meaningful constraints, demonstrating that even a single sightline can provide valuable information about the underlying physical parameters with this method.

\section{Comparison of inference methods}
\label{sec:comparison}

The first column of Fig.~\ref{fig:posterior-plots-all} most clearly shows the progressive improvement in inference accuracy using the five different methods. Importantly, this column corresponds to the least sensitive telescope configuration (only 50~hr with the uGMRT). As we go down this column from the top row to the bottom row, the effectiveness score, $E$, shows a strong improvement from 0.54 to 4.06.
As we go from Method~A2 to Method~B1, we notice that the use of XGBoost regression on the noisy power spectrum (Method~B1) instead of Bayesian MCMC already leads to a significant improvement in the inference accuracy in the part of the parameter space with $\mathrm{log}_{10}f_{\rm X} < -1$. 
When we further incorporate U-Net denoising (Method~B2), the inference accuracy improves further, even at $\mathrm{log}_{10}f_{\rm X} \approx -1$. 
Finally, Method~B3, which uses latent features from the U-Net encoder, achieves the best performance across all test points, demonstrating the power of this approach in extracting information from the data.

The part of the parameter space where all five methods struggle is the high-temperature high-ionization quadrant of the parameter space, represented by the test point $[\langle x_{\rm HI}\rangle^{\rm true}, \mathrm{log}_{10}f_{\rm X}^{\rm true}] = [0.11, -1]$. This is understandable as the signal is intrinsically weak for these parameter values. We discuss this in greater detail in Appendix~\ref{sec:difficult-parameter-space}. 

Fig.~\ref{fig:inference-multi-method} summarises the central result of this study by visually comparing the performance of different parameter inference methods through the size and shape of posterior contours for the test case $[\langle x_{\rm HI}\rangle^{\rm true}, \mathrm{log}_{10}f_{\rm X}^{\rm true}] = [0.11, -3]$. This point is part of the parameter space ($\langle x_{\rm HI}\rangle \lesssim 0.38$) that is preferred by the Ly$\alpha$ forest data (\citealt{2015MNRAS.447..499M}, \citealt{2022ApJ...932...76Z}, \citealt{2023MNRAS.525.4093G}, \citealt{2024ApJ...969..162D}, \citealt{2024MNRAS.530.3208G}), shown as grey shaded region in the figure.  We clearly see the improvement in the inference as we move from Method~A1 to Method~B3. The contours become progressively smaller and more accurately centred on the true parameter values, demonstrating the effectiveness of machine learning techniques in extracting information from noisy 21-cm forest data.

\section{Conclusions}
\label{sec:conclusion}

The 21-cm forest is a promising probe of the neutral intergalactic medium (IGM) during the epoch of reionization, offering a unique advantage over other 21-cm observables by being largely immune to foreground contamination. Nevertheless, its analysis remains challenging due to the intrinsic weakness of the signal and the prevalence of observational noise. In this work, we have systematically evaluated five inference techniques for extracting key physical parameters of the neutral IGM from noisy 21-cm forest data.

By constructing mock observations and developing machine learning models across a broad range of physical parameters, we have demonstrated the potential of hybrid deep learning pipelines to recover the properties of the neutral IGM at $z \sim 6$. Our analysis includes both scenarios with ten sightlines and the more challenging case of a single sightline. Building on previous studies that relied on the one-dimensional power spectrum of the 21-cm forest, we incorporate advanced machine learning methods to enhance inference accuracy and reduce the required integration times.

Specifically, we focus on constraining the mean neutral hydrogen fraction, $\langle x_{\mathrm{HI}}\rangle$, and the X-ray heating efficiency parameter, $\mathrm{log}_{10}f_{\rm X}$, during the epoch of reionization. Simulated datasets were generated using the semi-numerical code \textsc{21cmFAST}, with Gaussian noise added to mimic realistic observational conditions.

We have broadly considered two approaches to improve inference from noisy 21-cm forest data. The first is explicit noise subtraction, where the noise contribution is modelled (using the radiometer equation) and subtracted from the observed power spectra. In practice, this could also be achieved by measuring the noise directly from the data, for example, using spectral regions redward of the intrinsic 21-cm line in radio-loud quasar spectra. The second approach leverages machine learning to extract features from the data that are more informative than the power spectrum, without requiring explicit noise subtraction. In this work, we have systematically explored both strategies and find that the machine-learning-based feature extraction yields superior inference performance.

We evaluated the performance of both approaches across three telescope configurations: uGMRT over $50\,\rm hr$ and $500\,\rm hr$ integration times, and SKA1-low over $50\,\rm hr$ integration.  We quantified the inference performance using an effectiveness score, $E$, which measures the accuracy of the inferred parameters relative to their true values, probabilistically folding in the full posterior distribution. 

Out results are summarized in Table~\ref{tab:inference_comparison} and Fig.~\ref{fig:inference-multi-method}, which show our key findings:
\begin{itemize}
    \item Machine learning outperforms Bayesian inference.  XGBoost regression approach generally yields tighter constraints than MCMC. This demonstrates the strength of XGBoost in identifying complex non-linear relationships between the features (statistics of the observed spectrum) and the targets (physical quantities).
    \item Latent features from U-Net improve accuracy further. The best results are obtained using latent features extracted from the U-Net encoder, indicating that the model captures non-Gaussian information not encoded in the power spectrum.
    \item Machine learning reduces required integration time. For instance, the uGMRT over $50\,\rm hr$ dataset, with latent features achieves better performance ($E = 4.06$) than that of the $500\,\rm hr$ dataset with power spectra and Bayesian inference ($E = 2.67$), highlighting the potential of ML-based pipeline to help improve observational strategy. \
\end{itemize}

This work can be taken forward in several ways. 
First, the XGBoost regression model can be improved by incorporating uncertainty quantification techniques, such as quantile regression or Bayesian tree approaches, to better account for noise in the input data. This would help produce a more realistic predictive distribution that reflects true uncertainty, rather than just point estimates.
Second, while the U-Net architecture effectively denoises the 21-cm forest spectra, it does not fully capture the subtle spectral features, leading to over-subtraction. This can degrade inference accuracy, particularly in regions of the parameter space where the signal is weak. Future work could explore more advanced architectures or training strategies to enhance the U-Net's ability to recover these fine details. For example, incorporating attention mechanisms or multi-scale feature extraction could improve performance.
Third, one could use systematic hyperparameter optimization using tools such as Optuna \citep{2019arXiv190710902A} to refine the model.

We also note a few caveats regarding our modelling framework. First, we neglect the contribution of minihalos to the 21-cm forest signal, which may be non-negligible at certain scales. Accurately capturing their impact would require simulations with a larger dynamic range to resolve small-scale IGM structures. Second, our modeling framework does not currently account for radio frequency interference (RFI), which can significantly affect real observations. Incorporating realistic RFI models into our simulations and inference pipeline would be a potentially important direction for future work.

In conclusion, our study demonstrates the promise of hybrid deep learning pipelines in extracting astrophysical information from noisy 21-cm forest data. By significantly enhancing inference accuracy and reducing reliance on long integration times, such techniques could make it possible to probe the neutral IGM with currently operational facilities such as the uGMRT. 

We also propose that the usefulness of the techniques designed in this work are not limited to the 21-cm forest. The idea of XGBoost regression on latent-space U-Net encoding could provide a more accurate inference compared to traditional Bayesian MCMC techniques in a variety of situations such as the analysis of 21-cm intensity mapping data from the epoch of reionization, Ly$\alpha$ forest spectra, and quasar damping-wing absorption spectra, while requiring less intensive training requirements compared to other deep neural network approaches.

\section*{Acknowledgements}

It is a pleasure to thank Vipul Arora, Jatin Batra, James Bolton, Fred Davies, Abhirup Datta, Nirupam Roy, Shriharsh Tendulkar, and Nithyanandan Thyagarajan for helpful discussions. 
TŠ is supported by the Istituto Nazionale di Astrofisica Osservatorio Astronomico di Trieste (INAF-OATs) under the Theory grant `Cosmological Investigation of the Cosmic Web' (C93C23006820005) and by the Istituto Nazionale di Fisica Nucleare (INFN) INDARK grant. 
GK is supported by the Department of Atomic Energy (Government of India) research project with Project Identification Number RTI 4002. 
The authors acknowledge the computational resources provided by the Department of Theoretical Physics, Tata Institute of Fundamental Research (TIFR). 
This study benefitted from the following publicly available codes: \textsc{21cmFAST} \citep{Mesinger_2011_21CMFAST,Murray_2020}, \textsc{Astropy} \citep{Robitaille_2013}, \textsc{corner.py} \citep{ForemanMackey_2016_corner}, \texttt{emcee} \citep{Foreman-Mackey_2013_emcee}, \textsc{Matplotlib} \citep{Hunter_2007},  \textsc{NumPy} \citep{Harris_2020}, \textsc{SciPy} \citep{Virtanen_2020}, \textsc{PyTorch} \citep{paszke2019pytorch}, \textsc{XGBoost} \citep{chen2016xgboost} and \textsc{scikit-learn} \citep{2011JMLR...12.2825P}. We thank their developers for making these codes publicly available. 

\section*{Data Availability}

All code used in this work is available at \url{https://github.com/tomassoltinsky/21cmforest_1DPS}. This repository also includes the archived trained U-Net model and the XGBoost regressor used in this work. The simulation data used can be requested from the first author.

\bibliographystyle{mnras}
\bibliography{ml_infer_21cm_forest} 

\appendix 

\section{Likelihood test}\label{sec:likelihood_test}

\begin{figure}
    \begin{minipage}{1\columnwidth}
 	  \centering
 	  \includegraphics[width=\linewidth]{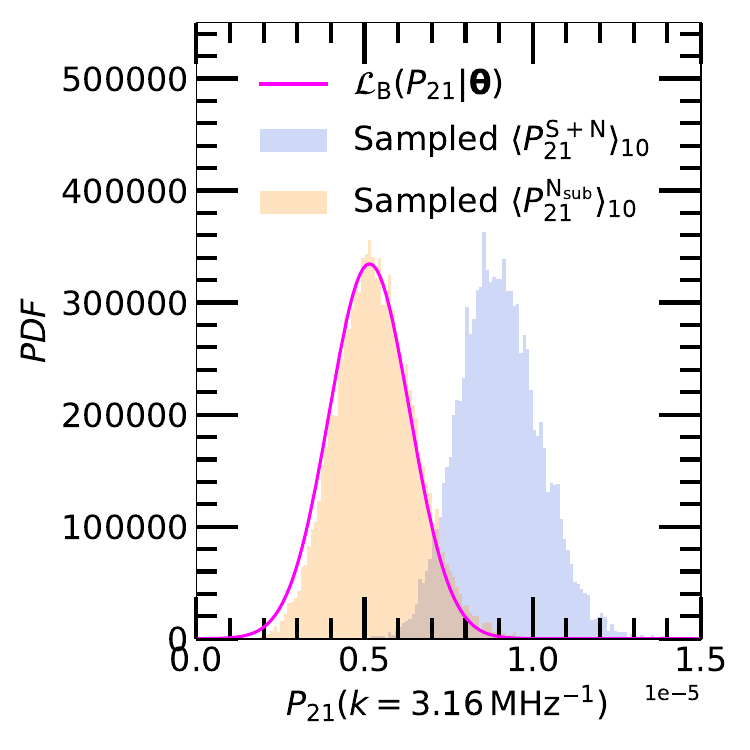}
 	  \includegraphics[width=\linewidth]{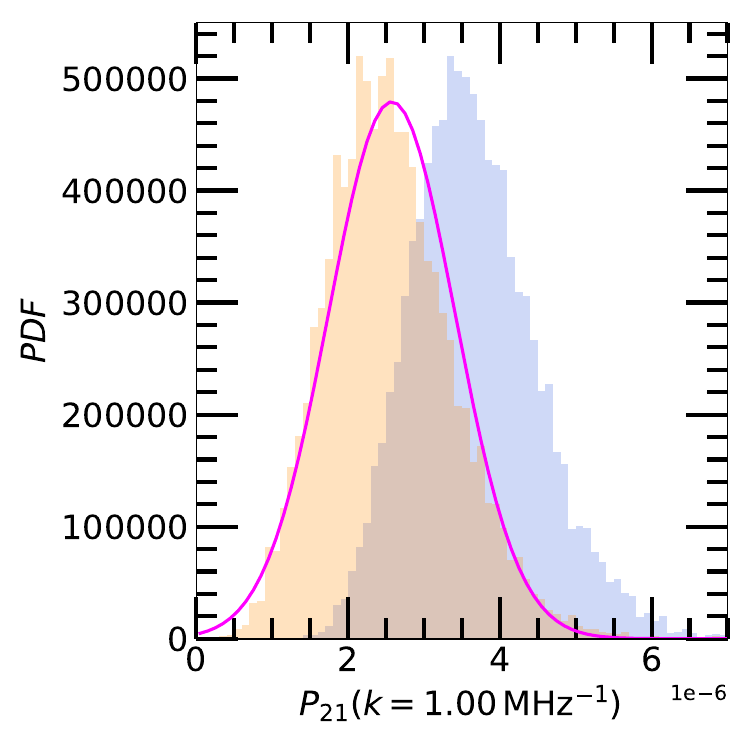}
    \end{minipage}
    \vspace{-0.3cm}
    \caption{Distribution of mock $\langle P_{21}^{\rm mock}\rangle_{10}$ at $k=3.16\,\rm MHz^{-1}$ (top panel) and $1.00\,\rm MHz^{-1}$ (bottom panel) for the model of $[\langle x_{\rm HI}\rangle^{\rm true}, \mathrm{log}_{10}f_{\rm X}^{\rm true}]=[0.52, -2]$ assuming a $50\,\rm hr$ per source observation by the uGMRT. Shown are two different cases for the mock observations, particularly the power spectrum which consists of both the signal and noise ($\langle P_{21}^{\rm S+N}\rangle_{10}$, blue) and noise-subtracted power spectrum ($\langle P_{21}^{\rm N_{sub}}\rangle_{10}$, orange). For comparison, the Gaussian likelihood used for the Bayesian statistics based inference, defined in equation (\ref{eq:likelihood_covar}), is indicated by the pink curve. Note different range of the horizontal axis between the panels.}
    \label{fig:likelihood_test}
\end{figure}

Here we test the validity of our choice of a Gaussian likelihood for the Bayesian inference analysis as defined in equation~(\ref{eq:likelihood_covar}). For this test, we use models with $\langle x_{\rm HI}\rangle^{\rm true}=0.52$ and $\mathrm{log}_{10}f_{\rm X}^{\rm true}=-2$ (green in Fig.~\ref{fig:posterior-plots-all}), and mock observations from the uGMRT with $t_{\rm int}=50\,\rm hr$ per source (corresponding to the left panels in Fig.~\ref{fig:posterior-plots-all}). In Fig.~\ref{fig:likelihood_test}, we show the likelihood for a particular $k$ (using only the relevant element of the covariance matrix and $\mathrm{\textbf{d}}$ for that $k$) as a pink curve. The top panel corresponds to $k=3.16\,\rm MHz^{-1}$, while the bottom panel shows $k=1.00\,\rm MHz^{-1}$. We also plot the distributions of mock power spectra including both signal and noise, $\langle P_{21}^{\rm S+N}\rangle_{10}$ (blue), and the noise-subtracted power spectra, $\langle P_{21}^{\rm N_{sub}}\rangle_{10}$ (orange), as described in Sec.~\ref{sec:MCMC_noisesubtracted}. The former corresponds to the analysis resulting in the posterior distributions in the top row of Fig.~\ref{fig:posterior-plots-all}, i.e. Method A1. It is evident that the distribution of mock power spectra deviates significantly from the assumed Gaussian likelihood: the presence of noise boosts the power, shifting the distribution to higher values. As a result, the MCMC inference tends to favor models with higher $P_{21}$ (i.e., more neutral and colder IGM), causing the posterior distributions to shift toward higher $\langle x_{\rm HI}\rangle$ and/or lower $\mathrm{log}_{10}f_{\rm X}$. In contrast, the noise-subtracted power spectrum $\langle P_{21}^{\rm N_{sub}}\rangle_{10}$ used in Method A2 is much better described by the Gaussian likelihood, leading to more accurate posterior distributions as shown in the second row of Fig.~\ref{fig:posterior-plots-all}.  At lower $k$ (comparing the top and bottom panels of Fig.~\ref{fig:likelihood_test}), the distributions overlap more closely, reflecting the increasing signal-to-noise ratio at these scales. However, even for the noise-subtracted case, the distribution is not perfectly Gaussian, and the Gaussian likelihood remains an approximation \citep{Wolfson_2023}.

\section{Feature importance}\label{sec:feature-imp}

\begin{figure}
    \centering
    \includegraphics[width=1\linewidth]{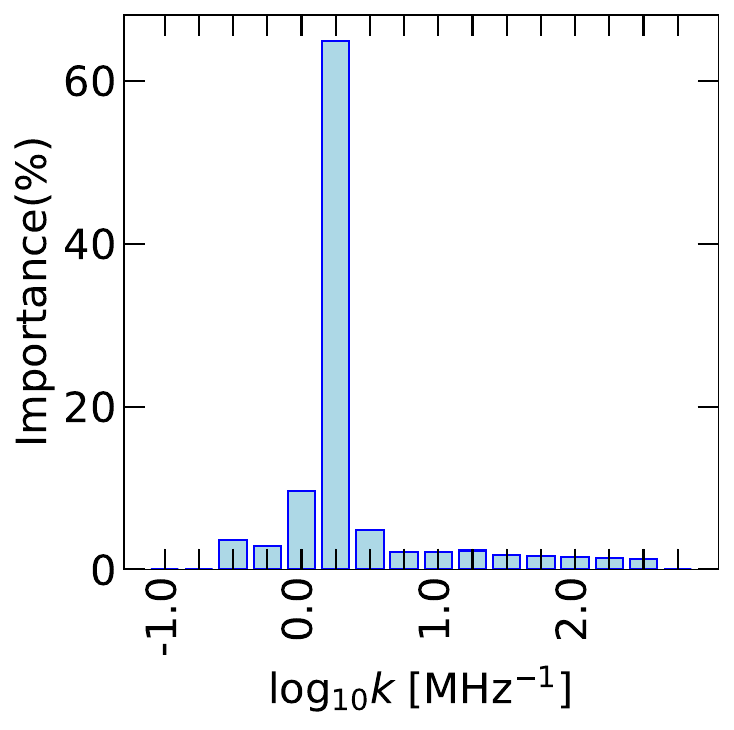}
    \caption{Feature importance scores assigned by XGBoost to each $k$-bin of the power spectrum. Higher bars indicate $k$-bins that contribute more strongly to the regression, highlighting the spectral scales most informative for parameter inference.}
    \label{fig:feature-imp}
\end{figure}

XGBoost estimates feature importance by tracking how often each input feature is used to split data across all trees in the ensemble. This frequency serves as a proxy for the feature’s contribution to the model's predictive performance. Fig.~\ref{fig:feature-imp} presents the importance scores assigned by XGBoost to various $k$-bins of the power spectrum during parameter inference. Notably, the $k$-bin centred at $k=1.78\,\mathrm{MHz}^{-1}$ emerges as the most influential feature in the models trained for both Method~B1 and Method~B2. This result can be explained by observing Fig.~\ref{fig:difficult-space-ps}. The power spectra show clearer separation in the range $1  \mathrm{MHz}^{-1} \lesssim k \lesssim 3 \mathrm{MHz}^{-1}$. This also aligns with findings from Fig.~6 of \citet{Soltinsky_2025}, where the 21-cm signal is shown to dominate over thermal noise and exhibits reduced sample variance near this scale, enhancing its discriminative power.    

\section{Inference Challenges in the Hot, Ionised Regime}\label{sec:difficult-parameter-space}

\begin{figure*}
    \centering
    \includegraphics[width=1.0\linewidth]{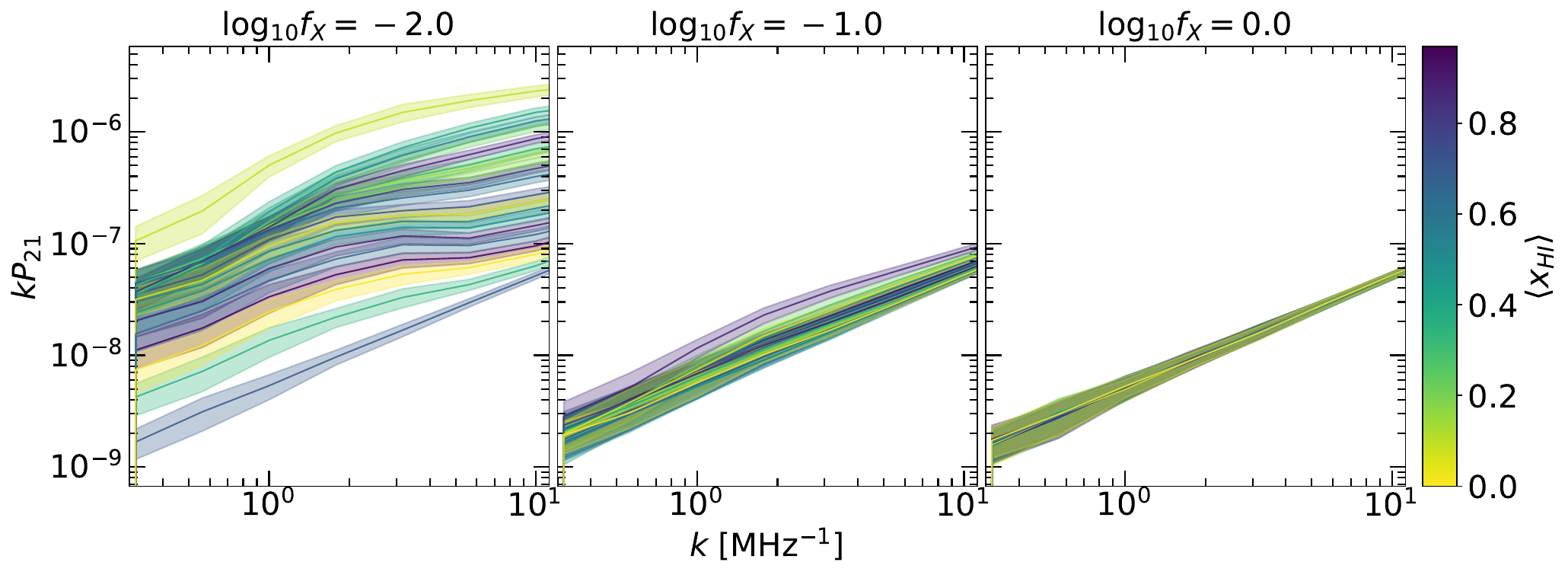}
    \caption{One-dimensional transmission power spectra of uGMRT $500\,\rm hr$ observations for different values of $\mathrm{log}_{10}f_{\rm X}$. Solid lines show the mean power spectra, while shaded regions indicate the $1\sigma$ sample variance. As $f_{\rm X}$ increases, the 21-cm signal weakens and thermal noise dominates, causing the power spectra for different $x_{\mathrm{HI}}$ values to overlap. This overlap reduces the ability to distinguish between models and limits inference accuracy in this regime.}
    \label{fig:difficult-space-ps}
\end{figure*}

Fig.~\ref{fig:difficult-space-ps} helps understand why it is difficult to infer the parameters in the region of the parameter space where $\mathrm{log}_{10}f_{\rm X}$ is large. The power spectra for different values of $x_{\mathrm{HI}}$ overlap significantly in this regime, making it challenging to distinguish between them. This is particularly problematic for the uGMRT over $50\,\rm hr$, where the noise dominates the signal, leading to a situation where the power spectra become nearly indistinguishable.  All of our methods struggle to accurately infer the parameters in this region, as the noise overwhelms the signal and the power spectra do not provide sufficient information to differentiate between the different $x_{\mathrm{HI}}$ values.  Physically, this is because the 21-cm absorption signal becomes too weak in this regime as the gas in the neutral IGM is too warm. 

\bsp
\label{lastpage}
\end{document}